\documentstyle{article}                      % [27.11.1995   bk95.tex]

\newcommand\quality{\textwidth 15.5cm \textheight 22cm} %\leftmargin 2cm

\quality 

\def\be{\begin{equation}}   \def\ee{\end{equation}}   \def\CR{$$ $$}
\def\IC{{\hbox{\rm C\kern-.58em{\raise.53ex\hbox{$\scriptscriptstyle|$}} 
    \kern-.55em{\raise.53ex\hbox{$\scriptscriptstyle|$}} }}}
\def\IN{\hbox{I\kern-.2em\hbox{N}}}
\def\IR{\hbox{\rm I\kern-.2em\hbox{\rm R}}}
\def\IZ{\hbox{{\rm Z}\kern-.3em{\rm Z}}}
\def\IT{\hbox{\rm T\kern-.38em{\raise.415ex\hbox{$\scriptstyle|$}} }}
\def\notsub{\hbox{$\subset$\kern-.55em\hbox{/}}}

\def\n{\noindent}                  \def\txt#1{{\rm #1}}
\def\var{\, {\rm var}}     \def\supp{\, {\rm supp}}
\def\const{\, {\rm Const} \,}                    \def\diam{\, {\rm diam}}
\def\dist{\, {\rm dist}}                         \def\mod1{\,({\rm mod\ } 1)\,}
\def\ep{\varepsilon}     \def\phi{\varphi}       \def\la{\lambda}
           
\def\1#1{\hbox{\large \bf 1}_{#1}}     % unit function on #1
\def\I|#1{{\big|}_{#1}}                % restriction to #1

\def\epd#1{\if #11 \ep^{-d} \else \ep^{-#1d} \fi}  % \epilon^{-#1d}
\def\proof{\smallskip \noindent {\bf Proof. \ }}
\newcommand\filledsquare{\ \vrule width 1.5ex height 1.2ex}  %filled square
\def\qed{\hfill\filledsquare\linebreak\smallskip\par}

\newcommand\Lpar{\left(}                \newcommand\Rpar{\right)}

\newtheorem{theorem}{Theorem}[section]   %Numbering: Theorem--Other section
\newtheorem{lemma}{Lemma}[section]       %{lemma}[theorem]{Lemma}   section
\newtheorem{proposition}[lemma]{Proposition}  %[theorem]
\newtheorem{corollary}[lemma]{Corollary}      %[theorem]
\newtheorem{definition}[lemma]{Definition}    %[theorem]
             \catcode`@=11 \@addtoreset{equation}{section} \catcode`@=12
 \newcommand{\ED}{{ED}} 
\newcommand{\BV}{{\rm BV}}     \newcommand{\bv}[1]{\|#1\|_{{\rm BV}}}
\newcommand{\Test}{{\cal T}_1} \newcommand{\ff}{{\tilde f}}
\newcommand{\TP}{{\rm TP}}    % Turning points
\newcommand{\PTP}{{\rm PTP}}  % Periodic turning points
\newcommand{\hyp}{4}  % hyperbolicity constant
\newcommand{\pbox}{\parbox{13.5cm}} \newcommand{\spread}{{\rm spread}}

\newcommand{\nz}{{\IN}}    \newcommand{\rz}{{\IR}}
  
\newcommand{\Id}{{\rm Id}}\newcommand{\F}{{\cal F}} \newcommand{\Z}{{\cal Z}}
\newcommand{\ph}{\varphi} \newcommand{\limn}{\lim_{n\to\infty}}

\newcommand{\liminfn}{\liminf_{n\to\infty}}

\newcommand{\beqa}{\begin{eqnarray}}   \newcommand{\eeqa}{\end{eqnarray}}
\newcommand{\beqas}{\begin{eqnarray*}} \newcommand{\eeqas}{\end{eqnarray*}}

\newcommand\mlbscale{1pt} %to change the scale: \renewcommand\mlbscale{1.3pt}

\def\bfig(#1,#2)#3#4{\begin{figure} \begin{center}
    \framebox{\setlength{\unitlength}{\mlbscale} \begin{picture}(#1,#2) #3 
    \end{picture} } \end{center} \caption{#4} \end{figure}}
\def\Bfig(#1,#2)#3#4{\begin{figure} \begin{center}
    \setlength{\unitlength}{\mlbscale} \begin{picture}(#1,#2) #3
    \end{picture} \end{center} \caption{#4} \end{figure}}
\def\bpic(#1,#2)#3{\setlength{\unitlength}{\mlbscale} 
    \begin{picture}(#1,#2) #3 \end{picture}} 
\makeatletter \newcount\@x@diff \newcount\@y@diff \newdimen\x@diff
\newdimen\y@diff \newcount\num@segments \newcount\num@segmentsi
\newif\if@flippedargs
\def\lineslope(#1,#2){
 \ifdim #1 <0pt \@xdim= -#1 \else\@xdim=#1\fi
 \ifdim #2 <0pt \@ydim= -#2 \else\@ydim=#2\fi
 \ifdim\@xdim >\@ydim \@tempdima=\@xdim \@xdim=\@ydim \@ydim=\@tempdima
 \@flippedargstrue\else\@flippedargsfalse\fi% x < y
 \ifdim\@ydim >1pt \@tempcnta=\@ydim
             \divide\@tempcnta by 65536% now \@tempcnta=integral part of #1.
             \divide\@xdim \@tempcnta\fi
 \ifdim\@xdim <.083333pt \@xarg=1 \@yarg=0
  \else\ifdim\@xdim <.183333pt    \@xarg=6 \@yarg=1
  \else\ifdim\@xdim <.225pt       \@xarg=5 \@yarg=1
  \else\ifdim\@xdim <.291666pt    \@xarg=4 \@yarg=1
  \else\ifdim\@xdim <.366666pt    \@xarg=3 \@yarg=1
  \else\ifdim\@xdim <.45pt        \@xarg=5 \@yarg=2
  \else\ifdim\@xdim <.55pt        \@xarg=2 \@yarg=1
  \else\ifdim\@xdim <.633333pt    \@xarg=5 \@yarg=3
  \else\ifdim\@xdim <.708333pt    \@xarg=3 \@yarg=2
  \else\ifdim\@xdim <.775pt       \@xarg=4 \@yarg=3
  \else\ifdim\@xdim <.816666pt    \@xarg=5 \@yarg=4
  \else\ifdim\@xdim <.916666pt    \@xarg=6 \@yarg=5
       \else                      \@xarg=1 \@yarg=1%
 \fi\fi\fi\fi\fi\fi\fi\fi\fi\fi\fi\fi
 \if@flippedargs\relax\else\@tempcnta=\@xarg \@xarg=\@yarg
                           \@yarg=\@tempcnta\fi
 \ifdim #1 <0pt \@xarg= -\@xarg\fi  \ifdim #2 <0pt \@yarg= -\@yarg\fi
}

\newif\if@toosmall \newif\if@drawit \newif\if@horvline \def\drawlinestretch{0} 
\def\drawline{\@ifnextchar [{\@idrawline}{\@idrawline[\drawlinestretch]}}
\def\@idrawline[#1](#2,#3){\@ifnextchar ({\@iidrawline[#1](#2,#3)}{\relax}}
\def\@iidrawline[#1](#2,#3)(#4,#5){\@drawline[#1](#2,#3)(#4,#5)
\@idrawline[#1](#4,#5)}

\def\@drawline[#1](#2,#3)(#4,#5){{%
\x@diff=#4\unitlength \advance\x@diff by -#2\unitlength
\y@diff=#5\unitlength \advance\y@diff by -#3\unitlength

\ifx\@linefnt\tenln \linethickness{0.5pt} \else \linethickness{0.9pt}\fi
\lineslope(\x@diff,\y@diff)% returns the two integers in \@xarg & \@yarg.

\@toosmalltrue
{\ifdim\x@diff <\z@ \x@diff=-\x@diff\fi
 \ifdim\y@diff <\z@ \y@diff=-\y@diff\fi
 \ifdim\x@diff >10pt \global\@toosmallfalse\fi
 \ifdim\y@diff >10pt \global\@toosmallfalse\fi}

\@drawitfalse\@horvlinefalse \ifnum#1 <0 \relax\else\@horvlinetrue\fi
\if@toosmall\@horvlinetrue\fi 
\if@horvline
 \ifdim\x@diff =0pt \put(#2,#3){\ifdim\y@diff >0pt \@linelen=\y@diff \@upline
                                 \else\@linelen=-\y@diff \@downline\fi}%
 \else\ifdim\y@diff =0pt
          \ifdim\x@diff >0pt \put(#2,#3){\vrule \@height \@halfwidth \@depth
                                \@halfwidth \@width \x@diff}
                \else \put(#4,#5){\vrule \@height \@halfwidth \@depth
                                \@halfwidth \@width -\x@diff}\fi
       \else\@drawittrue\fi\fi % construct the line explicitly
\else\@drawittrue\fi

\if@drawit
\ifnum\@xarg< 0 \@negargtrue\else\@negargfalse\fi
\ifnum\@xarg =0 \setbox\@linechar%
\hbox{\hskip -\@halfwidth \vrule \@width \@wholewidth \@height 10.2pt
 \@depth \z@}
\else \ifnum\@yarg =0 \setbox\@linechar%
\hbox{\vrule \@height \@halfwidth \@depth \@halfwidth \@width 10.2pt}
\else \if@negarg \@xarg -\@xarg \@yyarg -\@yarg
      \else \@yyarg \@yarg\fi
\ifnum\@yyarg >0 \@tempcnta\@yyarg \else \@tempcnta -\@yyarg\fi
\setbox\@linechar\hbox{\@linefnt\@getlinechar(\@xarg,\@yyarg)}%
\fi\fi

\if@toosmall % => it isn't a horiz or vert line and is toosmall.
  \@dottedline[\picsquare]{.98\@wholewidth}%
(#2\unitlength,#3\unitlength)(#4\unitlength,#5\unitlength)%
\else

\ifnum\@xarg=0\relax\else\ifdim\x@diff >\z@ \advance\x@diff -\wd\@linechar
  \else\advance\x@diff \wd\@linechar\fi\fi
\ifnum\@yarg=0\relax\else\ifdim\y@diff >\z@\advance\y@diff -\ht\@linechar
  \else\advance\y@diff \ht\@linechar\fi\fi
\ifdim\x@diff <\z@ \@x@diff=-\x@diff \else\@x@diff=\x@diff\fi
\ifdim\y@diff <\z@ \@y@diff=-\y@diff \else\@y@diff=\y@diff\fi
\num@segments=0 \num@segmentsi=0
\ifdim\wd\@linechar >1pt
 \num@segmentsi=\@x@diff \divide\num@segmentsi \wd\@linechar\fi
\ifdim\ht\@linechar >1pt
 \num@segments=\@y@diff \divide\num@segments \ht\@linechar\fi
\ifnum\num@segmentsi >\num@segments \num@segments=\num@segmentsi\fi
\advance\num@segments \@ne %to account for round-off error
\ifnum #1=0 \relax \else\ifnum #1 < -99
  \typeout{***drawline: reduction <= -100 percent implies blankness!***}
\else\num@segmentsi=#1 \advance\num@segmentsi by 100
     \multiply\num@segments \num@segmentsi
     \divide\num@segments by 100
\fi\fi
\divide\x@diff \num@segments
\divide\y@diff \num@segments
\advance\num@segments \@ne %for the last segment for which I subtracted
\@xdim=#2\unitlength \@ydim=#3\unitlength
\if@negarg \advance\@xdim -\wd\@linechar\fi
\ifnum\@yarg <0 \advance\@ydim -\ht\@linechar\fi

\@killglue
\loop \ifnum\num@segments > 0
      \unskip\raise\@ydim\hbox to\z@{\hskip\@xdim \copy\@linechar\hss}%
      \advance\num@segments \m@ne\advance\@xdim\x@diff\advance\@ydim\y@diff%
\repeat \ignorespaces 
\fi%the if of @toosmall
\fi}}% for \if@drawit
\newcounter{@sc} \newcounter{@scp} \newcounter{@t} \newlength{\@x}
\newlength{\@xa} \newlength{\@xb}  \newlength{\@y} \newlength{\@ya}
\newlength{\@yb} \newsavebox{\@pt}
\def\bezier#1(#2,#3)(#4,#5)(#6,#7){\c@@sc#1\relax
 \c@@scp\c@@sc \advance\c@@scp\@ne
 \@xb #4\unitlength \advance\@xb -#2\unitlength \multiply\@xb \tw@
 \@xa #6\unitlength \advance\@xa -#2\unitlength
 \advance\@xa -\@xb \divide\@xa\c@@sc
 \@yb #5\unitlength \advance\@yb -#3\unitlength \multiply\@yb \tw@
 \@ya #7\unitlength \advance\@ya -#3\unitlength
 \advance\@ya -\@yb \divide\@ya\c@@sc
 \setbox\@pt\hbox{\vrule height\@halfwidth depth\@halfwidth
 width\@wholewidth}\c@@t\z@
 \put(#2,#3){\@whilenum{\c@@t<\c@@scp}\do
 {\@x\c@@t\@xa \advance\@x\@xb \divide\@x\c@@sc \multiply\@x\c@@t
 \@y\c@@t\@ya \advance\@y\@yb \divide\@y\c@@sc \multiply\@y\c@@t
 \raise \@y \hbox to \z@{\hskip \@x\unhcopy\@pt\hss}\advance\c@@t\@ne}}}
\begin{document} 
\newenvironment {bem} {\begin{bemerkung} \rm }{\end{bemerkung}}
\newtheorem {bemerkung}{Remark}[section]
\def\CP{{\cal P}}

\title{Stochastic stability versus localization 
       \\ in chaotic dynamical systems}
\author{Michael Blank
         \thanks{On leave from Russian Academy of Sciences, Inst. for 
         Information Transmission Problems, 
         Ermolovoy Str. 19, 101447, Moscow, Russia, blank@obs-nice.fr}
        , Gerhard Keller
        \\ \\ 
         Mathematisches Institut, Universitat Erlangen-Nurnberg \\
         Bismarckstrasse 1 1/2, D-91054 Erlangen, Germany.
        }
\date{\today}
\maketitle
\n {\bf Abstract} -- We prove stochastic stability of chaotic maps for a 
general class of Markov random perturbations (including singular ones) 
satisfying some kind of mixing conditions. One of the consequences of this 
statement is the proof of Ulam's conjecture about the approximation 
of the dynamics of a chaotic system by a finite state Markov chain. 
Conditions under which the localization phenomenon (i.e. stabilization of 
singular invariant measures) takes place are also considered. 
Our main tools are the so called bounded variation 
approach combined with the ergodic theorem of Ionescu-Tulcea and Marinescu,
and a random walk argument that we apply to prove the absence of ``traps'' 
under the action of random perturbations. \bigskip

\section{Introduction} \label{Int}

The investigation of stochastic stability of chaotic dynamical systems has a
long history, and in this paper we shall restrict ourselves
mainly to the case of one-dimensional piecewise expanding maps (i.e.
Lasota-Yorke \cite{LY} type maps). It is well known that if the expanding
constant is larger than $2$ a Lasota-Yorke type map is stable with
respect to mixing random perturbations. It is also known that the
assumption that the expanding constant is larger than $1$ is not enough
for stochastic stability. The point is, only when the expanding
constant is larger than $2$, the map is really locally expanding, i.e.
the image of any small enough interval has Lebesgue measure larger than
the genuine one. We prove that for a very general class of Markov
random perturbations the  the mixing condition on perturbations is
sufficient for stochastic stability. One of the consequences of this
statement is the proof of so called Ulam's conjecture, about the
approximation of a chaotic system by a finite state Markov chain.
Conditions under which the localization phenomena (i.e. stabilization
of singular invariant measures) takes place are also considered. The
nature of this phenomena is a mixing of unstable and stable directions
under the action of arbitrary small perturbations. To show that the
localization is not the result of discontinuities in the considered
class of maps we prove the localization under the action of small
mixing random perturbations for general smooth hyperbolic
$d$-dimensional maps.

In this paper we discuss some unusual phenomena that appear in 
chaotic dynamical systems. First of all, what does it mean a {\em chaotic} 
system or map. We shall consider only discrete time {\em dynamical 
systems}, i.e. pairs $(f,X)$, where $X \subset \IR^d$ is a compact phase 
space (say $X=[0,1]^d$) and $f:X \to X$ is a nonsingular map, iterations 
of which define trajectories of the dynamical system. Nonsingular 
means that $m(f^{-1}A)>0$ for any measurable set $A \subseteq X$ with 
positive Lebesgue measure $m(A)>0$.

We shall be interested mainly in asymptotic, when time tends to
infinity, properties of systems under consideration.

\smallskip 
Consider now small {\em random perturbations} of the discrete time 
dynamical system. Before giving a rigorous definition of random 
perturbations of dynamical systems we mention one example, which will 
be useful to have in mind later on. Let $U \subset R^d$ be a 
neighborhood of a compact invariant set of the map $f$ such that $fU 
\subset U$. Suppose that, when we apply $f$ to a point $x \in U$, rather than 
choosing the exact value of $fx$ we choose in a random way, in 
accordance with a homogeneous distribution, a point from the ball 
$B_\ep(fx)$ (i.e. from the ball with centre at the point $fx$ and 
radius $\ep$). The resulting random map corresponds to independent 
homogeneously distributed random perturbations.
\begin{definition} Let $Q_\ep(x,A)$ be a family of transition probabilities
and $f:X\to X$ a map. We denote by $f_\ep$ the Markov 
process on the phase space $X$, defined by transition the probabilities 
$Q_\ep(fx,A)$ and call $f_\ep$ a random perturbation of $f$.
\end{definition}
We are mainly interested in {\em small} perturbations, small in the sense that
\be\label{small-ass}
\|Q_\ep h-h\|_1\to 0\quad\mbox {as $\ep\to 0$ for any $h\in L^1$,}
\ee
where $\|\,.\,\|_1$ denotes the $L^1$-norm.
Additionaly we assume that our perturbations are
{\em local}, i.e.
\be\label{local-cond}
\mbox{
$Q_\ep(x,A) = 0$ for any pair $(x,A)$, such 
that $\dist(x,A) > \ep$.
}
\ee
Therefore the parameter $\ep$ here plays the
role of a ``magnitude'' of the perturbation. However, this 
local condition does not imply 
stochastic stability even for very smooth chaotic maps as we show in 
Lemma \ref{lemma2.1}. Thus to 
obtain some kind of stochastic stability the perturbations should 
satisfy at least some kind of  mixing condition (say, for example, of Doeblin's 
type \cite{Do}). One of the most general assumptions of this type for random 
perturbations was proposed in \cite{Bl18}: 
\be 
\var(Q_\ep h) \le \var(h) + C\|h\|_1 \label{var-as} 
\quad\mbox{(regularity assumption)}
\ee
for any function $h$ of bounded variation ($\var(h) < \infty$). 
Remark, that for independent random perturbations 
the transition density depends only on the difference between its 
arguments: $q_\ep(x,y)=q_\ep(x-y)$, and the constant $C$ on the right hand 
side of the inequality~(\ref{var-as}) may be set to zero. Thus this condition 
means that the perturbation is not too far from the independent one. 
Besides to convolutions with smooth transition probabilities our results apply to
many other types of perturbations like bistochastic smooth perturbations,
singular perturbations of point mass type and deterministic perturbations
by chaotic maps close to identity.
\par
The basic idea of our approach is the following:
Consider the Perron-Frobenius operator $P$, describing the 
dynamics of densities under the action of the map $f$, on 
Banach space $(\BV,\bv{\,.\,})$ of functions of bounded variation.
Here $\bv{h}=\var(h)+\|h\|_1$.
We prove that the transition operator $P_\ep = Q_\ep P$ of the randomly perturbed 
map satisfies the uniform Lasota-Yorke type inequality
\be\label{LY-type}
\bv{P_\ep^N h}
\le
\alpha\cdot\bv{h}+C\cdot\|h\|_1 \quad (h\in\BV)
\ee
for some fixed integer $N$, $\alpha\in(0,1)$ and $C>0$ independent of $\ep$.
This yields at once the existence of $f_\ep$-invariant densities $h_\ep$
with $\bv{h_\ep}\le\frac C{1-\alpha}$, such that (\ref{small-ass}) forces
each limit $h_*=L^1-\lim_{\ep\to 0}h_\ep$ to be an invariant density for $f$.
If the expanding constant $\lambda$ is larger than $2$, (\ref{LY-type}) was proved with $N=1$ in various settings, see {\em e.g.}
\cite{{Ke},{Bl5},{Ki}} and 
references therein.
For quite a while it was supposed that the extension of this inequality
to cases with $\lambda\in(1,2]$ is only a
technical problem. However, the counterexample constructed in
\cite{Bl17} shows that the situation is not so simple.
After this counterexample (which we shall discuss in Section~\ref{counter1}) it
became clear, that the main problem is the possible existence of periodic
{\em turning} points, {\em i.e.} the points, 
where the derivative of the map $f$ 
is not well defined. Namely, 
under the action of random perturbations ``traps'' or 
``absorbing sets'' can appear near these periodic turning points, 
which leads to the appearance of new ergodic 
components in the perturbed system. In \cite{BY} 
stochastic 
stability was proved for convolution type smooth random perturbations in  
situations with $\lambda\in(1,2]$ but without periodic turning points. 
However it is not clear whether their argument is works for general Markov 
perturbations.
\par
The main result obtained in these papers is the stability of statistical 
properties of the perturbed systems in the zero noise limit. We shall show 
that in the general case there may be generalized phase transitions 
of type localization - delocalization in such systems. These phase 
transitions correspond to situations, when trajectories, which 
should normally be dense, remain confined to a small region (which 
vanishes, when the coupling constant goes to zero). We call this 
``localization phenomenon''. The first observation of this type was 
published in \cite{Bl17}.  
\begin{definition} Let $X=[0,1]$. A map $f:X\to X$ is {\em piecewise expanding} 
(PE) if there exists a partition of $X$ into disjoint 
intervals $\{X_j\}$, such that $f\I|{\txt{Clos}(X_j)}$ is a 
$C^2$-diffeomorphism (from the closed interval $\txt{Clos}(X_j)$ to its 
image), if the {\em expanding constant} of the map
$$ \la_f := \inf_{j, \, x \in X_j} |f'(x)|, $$
is positive, and if $\la_{f^\kappa}>1$ for some iterate $f^\kappa$.
\end{definition}
For some of our results we impose some extra conditions on $f$. In particular we distinguish between
three different ``hyperbolicity'' 
assumptions:
\begin{itemize}
  \item $f \in H_\infty$, if $f$ is a general PE map;
  \item $f \in H_1$, if $f$ is a PE map with $\la_f>1$;
  \item $f \in H_\infty^\ep$ if $f$ is a PE map for which there is $\lambda>1$
      such that 
        $\prod_{k=1}^n|f'(f_\ep^k x)| \ge \la^n$
      for each $\ep$-trajectory $(f_\ep^kx)_{k=1,2,\ldots}$.
\end{itemize}
Clearly $H_1\subset H_\infty^\ep\subset H_\infty$.
\par
Typical examples of PE maps are shown in Figure~\ref{maps1}. Observe that 
maps that we consider need not to be continuous. As it is well known, 
starting from the paper of Lasota \& Yorke \cite{LY}, these maps have 
all the statistical properties that one can reasonably expect 
from deterministic 
dynamical system. They have a smooth (absolutely continuous) invariant 
measure $\mu_f$ (Sinai-Bowen-Ruelle measure of this map), exponential 
correlation decay and they obey a 
CLT with respect to this measure, etc \cite{HK,Bl5}. In this 
paper we want to discuss stability of these properties with respect to 
small random perturbations. We shall emphasize some aspects of this 
problem, because they seem quite counterintuitive, at least from our 
point of view.
\begin{definition} The {\em image} of a measure $\mu$ under the action of 
a map $f$ is the measure $f\mu$ defined by $f\mu(A)=\mu(f^{-1}A)$ for any 
measurable set $A$. By $f_\ep \mu$ we mean the measure
$f_\ep \mu(A)=\int Q_\ep(fx,A)d\mu(x)$. A measure $\mu$ is $f$\ ($f_\ep$)--invariant, if $f\mu=\mu$\ ($f_\ep\mu=\mu$). $\mu$ is called 
{\em smooth}, if it has a density with respect to Lebesgue measure.
\end{definition}

\begin{definition} 
A probability measure $\mu_f$ is called 
a {\em Sinai-Bowen-Ruelle (SBR)} measure of $f$, if
there exists an open subset $U$ of the phase space such 
that the images 
$f^n \mu$ 
of any smooth probability measure $\mu$ with the support in $U$ converge weakly to $\mu_f$. Analogously we define SBR measures $\mu_\ep$ for perturbed systems
$f_\ep$.
\end{definition}

\begin{definition} A {\em turning point} of a map $f$ is 
a point, where the derivative of the map is not well defined. The set 
of turning points is denoted by $\TP$, the set of periodic turning points by
$\PTP$.
(In Figure~\ref{maps1}a we have $c\in\TP$, in Figure~\ref{maps1}b we have
$c\in\PTP$.) 
\end{definition}
{\bf Standing assumption:} From now on we assume that all perturbations $Q_\ep$
satisfy (\ref{small-ass}), (\ref{local-cond}) and (\ref{var-as}).
%
%
%%%%%%%%%%%%%%%%%%%%%%%%%%%%%%%%%%%%%%%%%%%%%%%%%%%%%%%%%%%%%%%%%%
%% Chaotic maps
\Bfig(350,150)
    {% picture (a)
     \drawline(0,0)(150,0)(150,150)(0,150)(0,0)
     \drawline[-40](0,0)(150,150)
     \put(73,-10){$c$} \drawline[-40](75,0)(75,150)
     \put(113,75){$\la_f$} \put(70,155){{\bf(a)}}
     \put(0,-10){$0$} \put(150,-10){$1$} \put(-10,145){$1$}
     \thicklines \drawline(75,125)(150,20)
     \bezier{200}(0,0)(30,90)(75,140)
     % picture (b)
     \put(200,0){\bpic(150,150){\thinlines
     \drawline(0,0)(150,0)(150,150)(0,150)(0,0)
     \drawline[-50](0,0)(150,150) \drawline[-50](75,0)(75,75)
     \thicklines
     \drawline(0,150)(30,0)(75,75)(120,0)(150,150)
     \drawline[-20](40,26)(75,85)(110,26)
     \put(73,-10){$c$} \put(70,155){{\bf(b)}}
     \put(0,-10){$0$} \put(150,-10){$1$} \put(-10,145){$1$}
     }}
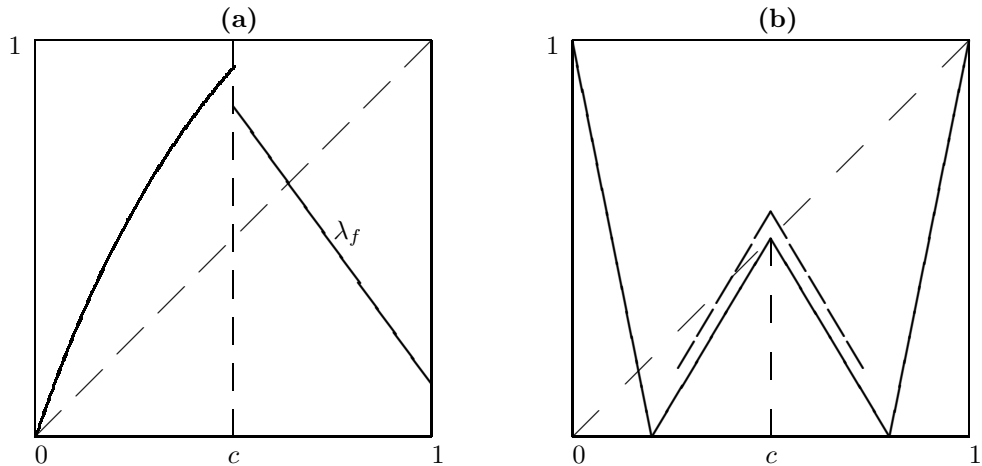
    }{PE maps. (a) general PE map, (b) W-map \label{maps1}}
%%%%%%%%%%%%%%%%%%%%%%%%%%%%%%%%%%%%%%%%%%%%%%%%%%%%%%%%%%%%%%%%%%
%
%
\begin{theorem} {\bf\cite{Bl17,Ke}} \label{stab1} 
Suppose $\la_f>2$ and that $f$ has a unique smooth SBR measure $\mu_f$. Then 
for any sufficiently small $\ep>0$ also the perturbed system $f_\ep$ has a
unique smooth 
invariant SBR measure $\mu_\ep$, and the $\mu_\ep$ converge weakly
to $\mu_f$ as $\ep \to 0$.  
\end{theorem}

In this paper we generalize
Theorem~\ref{stab1} and show that at least some additional assumptions 
are necessary for the stability of the smooth SBR measure $\mu_f$.

\begin{theorem} \label{stab2} Let $f \in H_\infty$, 
$\PTP=\emptyset$, and suppose that $f$ has a unique smooth
SBR measure $\mu_f$. Then for any $\ep>0$ small enough the perturbed 
system $f_\ep$ has a smooth invariant SBR measure $\mu_\ep$ 
converging weakly as $\ep \to 0$ to the smooth SBR measure $\mu_f$.
\end{theorem}

So the only topological obstacle for the stability of statistical 
properties of PE maps under small perturbations is the existence 
of periodic turning points. Consequently, generic 
PE maps are stochastically stable. However, in order to investigate 
stability properties of families of maps, rather than individual 
maps, also maps with periodic turning 
points must be studied.

\begin{theorem} \label{instab1} There exists a map $f \in H_1$ with 
periodic turning points and smooth perturbations $Q_\ep$
such that for sufficiently small $\ep>0$  
the perturbed system $f_\ep$  has a unique invariant SBR measure $\mu_\ep$,
converging to a singular and unstable invariant measure of the map $f$ 
as $\ep \to 0$.
\end{theorem}
                       
This theorem does not only state the localization of 
$f_\ep$--invariant measures 
(supports on small sets), but also guarantees their smoothness and 
the absence of other SBR measures. Actually the localization 
phenomenon was firstly shown in \cite{Bl17,Bl18}, but the statement 
about the smoothness of invariant measures, absence of other SBR 
measures, and investigation of their properties are new. 
\par
The nature of the localization phenomenon here is due to the fact that 
if the expanding constant $\la_f$ is less than $2$ the map $f$ is not 
locally expanding near periodic singular points. Indeed it may map 
a small interval $\Delta$ to the interval of length $\la_f |\Delta|/2$.
This property distinguishes
the situations in Theorems~\ref{stab1} and \ref{stab2}.
To show that this phenomenon is not something obscure, specific for only
discontinuous maps, we shall prove its presence for a general
multidimensional smooth hyperbolic map in Section~\ref{counter1}.
\par
In order to exclude the behaviour described in Theorem \ref{instab1} we introduce the following {\em random walk} ($RW$) assumption:
\be\label{RW-assumption}
\pbox{
A family $Q_\ep$ ($\ep>0$) belongs to the class $RW$, if there are 
$0<\theta,\delta<1$ such that
\centerline{
$Q(x,[0,x-\theta\ep])>\delta$ and
$Q(x,[x+\theta\ep,1])>\delta$ for all $x$.
}}
\ee
For the next two theorems we assume additionally that a local version of
(\ref{var-as}) is satisfied, see (\ref{QLYassumption}).
\begin{theorem} \label{stab3} 
Let $f \in H_\infty$, 
$\PTP$ not necessarily empty, and suppose that
the transition probabilities $Q_\ep$ belong to $RW$ and have 
densities $q_\ep(x,y)$ satisfying
\[
\var(y\mapsto q_\ep(x,y))\le\frac M\ep\quad\mbox{ for all $x$}\ .
\]
Then for any 
sufficiently small $\ep>0$ the perturbed 
system $f_\ep$ has a smooth invariant SBR measure $\mu_\ep$ converging 
weakly to the smooth SBR measure $\mu_f$ as $\ep \to 0$. 
\end{theorem}
To formulate results about more general singular perturbations we need 
the notion of renormaliziability, which will be disscussed in detail 
at the end of 
Section \ref{close}. Roughly speaking this notion means that for any $\ep>0$ 
small enough a map (or a perturbation) can be renormalized to a fixed shape
in small neighbourhoods (of order $\ep$) of periodic turning points.
\begin{theorem} \label{stab4} Let $f \in H_\infty^\ep$,
$\PTP$ not necessarily empty, and $Q \in RW$. Then
for any $\ep>0$ small enough the perturbed system $f_\ep$ has a smooth
invariant SBR measure $\mu_\ep$ and the transition operator $Q_\ep P_f$ 
satisfies the Lasota-Yorke type inequality. If additionally both the map 
and the perturbations are renormalizable, then the measures $\mu_\ep$ 
converge weakly as $\ep \to 0$ to the smooth SBR measure $\mu_f$.
\end{theorem}

The last result of the paper 
is the proof of Ulam's conjecture \cite{Ul} on the 
approximation of chaotic dynamics by finite state Markov chains. The idea
of the construction is to take a finite partition $\{\Delta_i\}$ of the 
phase space with bounded volume ratios
and to approximate the action of the map $f$ by the Markov 
chain with transition probabilities
$$ p_{ij}:=\frac{|\Delta_i \cap f^{-1}\Delta_j|}{|\Delta_i|}.$$
This construction could be considered as a special type of small random 
perturbations.
In fact, our standing assumption on $Q_\ep$ is satisfied for this particular class of perturbations.
The correctness of Ulam's conjecture was proved in 
\cite{Li} for PE maps with $\la_f>2$. In contrast to our other statements
the answer to the conjecture does not 
depend on the existence of periodic turning points.

\begin{theorem} \label{ulam} Let $f \in H_\infty$. Then invariant
measures constructed by Ulam's procedure converge weakly to
the smooth SBR measure $\mu_f$.
\end{theorem}
\section{Counterexamples} \label{counter1}
\subsection{Localization in the absence of the regularity assumption}
Our first example shows that even for arbitrarily smooth chaotic maps
and absolutely continuous perturbations
the {\em local} condition (\ref{local-cond}) alone
does not imply stochastic
stability .
\begin{lemma}\label{lemma2.1}
Suppose $f:X\to X$ has a cycle (periodic trajectory) 
$\bar c= c_1, \dots, c_n$ and satisfies a Lipschitz condition in
some neighborhood of this cycle. Then there exists a family of local
random perturbations with densities $q_\ep(x,y)$ for which
the $f_\ep$ have smooth invariant measures $\mu_\ep$
converging to the $\delta$-measure on this cycle when $\ep \to 0$.
\end{lemma}
\proof Fix a constant $\beta$ greater than the local Lipschitz constant 
of the map $f$ and consider a family of transition probability densities
$$q_\ep(x,y)=\cases{1/|B_{\ep/\beta}(\bar c)|,  &if $x \in B_\ep(\bar c)$ and 
                                        $y \in B_{\ep/\beta}(\bar c)$; \cr
                    0,              &if $x \in B_\ep(\bar c)$ and 
                                        $y \notin B_{\ep/\beta}(\bar c)$; \cr 
                    1/|B_\ep(\bar c)|, &if $x \notin B_\ep(\bar c)$ and 
                                        $y \in B_\ep(x)$; \cr
                    0,              &otherwise. \cr}  $$
Then clearly $f_\ep$ has a
new ergodic component in the $\ep$-neighborhood of the cycle $\bar c$,
because $fB_{\ep/\beta}(\bar c) \subset B_\ep(\bar c)$ and by the
construction $q_\ep(x,y)=0$ for $x\in B_\ep(\bar c)$ and 
$y\not\in B_{\ep/\beta}(\bar c)$. 
\qed
This proof shows that for stochastic stability some additional 
assumption is needed 
that prevents the perturbation to act precisely against the
dynamics of the map. This is one of the purposes of the regularity assumption (\ref{var-as}).
\subsection{Periodic turning points and the $RW$ assumption,
proof of Theorem~1.3.}
Fix a number $10<\beta<\infty$ and consider the following family of transition 
probability densities:
\be
q_\ep(x,y)=\cases{\frac{\beta}{2\ep}, &if $(1-\frac1\beta)\ep \le -x+y 
                                           \le (1+\frac1\beta)\ep$; \cr
                      0,                  &otherwise. \cr} \label{tr-den} 
\ee
Note that random perturbations with such transition densities are of 
convolution 
type and have a very strong mixing property - exponential rate of correlation 
decay. The following lemma proves at the same time Theorem \ref{instab1}:

\begin{lemma} [\cite{Bl18}] Let $1 < \la < 2-1/beta$, and let the map $f$ is 
defined as follows: 
\be f(x) = \cases{1 - \frac{2x}{1-1/\la}, &if $0 \le x \le (1-1/\la)/2$; \cr
                  \la x - \frac{\la-1}2,  &if $(1-1/\l)/2 < x \le 1/2$; \cr
                  f(1-x),                 &otherwise. \cr}  \label{w-map} \ee
See Figure~\ref{maps1}.b. 
Then for all sufficiently small $\ep>0$ the perturbed map
$f_\ep$ has a unique invariant measure converging to the $\delta$-measure 
at $\frac 12$
as $\ep \to 0$. 
\label{local-ex} \end{lemma}
The {\bf proof} of this statement is based on the fact 
that trajectories of $f_\ep$
starting from 
$[1/2 - \ep(\la(\beta+1)-\beta+1)/\beta, \, 1/2 + \ep(\beta+1)/\beta]$ remain 
in this interval for ever with probability $1$. A related example with the same map $f$ but different transition kernel was given in \cite{Ke}.
\\
We remark that, as $|f'(x)|=\la<2$
in some neighborhood of the fixed point $c \in X$, the mapping
$f$ is not expanding in some sense, because it maps an interval neighborhood
$U$ of the fixed point to the interval $fU$ where
$$         |fU| \le \frac{\la \, |U|}2 < |U|. $$
So, in the presence of periodic turning points, instability 
of these points under $f_\ep$ 
is a necessary extra requirement. For this reason we introduced the
random walk assumption $RW$.
\par
In a broader context the inequality $|fU|<|U|$ can
be interpreted as a  mixing of stable and 
unstable directions under the action of random perturbations.
In fact, a more general statement is:

\begin{theorem} Let a $C^2$-smooth map $f:\IR^d \to \IR^d$ has
a hyperbolic fixed point $c \in \IR^d$, such that for the linear
part $D_c f$ at this point we have
$$ \IR^d=E_c^s+E_c^u, \quad \dim(E_c^s) \ge \dim(E_c^u) = N ,$$
$$ |D_c f(x)| \le \la_s |x|  {\rm \ \ for all \ } x \in E_c^s ,$$
$$ \la_u |x| \le |D_c f(x)| \le \Lambda_u |x| 
   {\rm \ \ for all \ } x \in E_c^u ,$$ 
and $\la_s \Lambda_u < 1$. Then there exist a family of local mixing stochastic
$\ep$-perturbations such that for all $\ep >0$ small enough the
stochastically $\ep$-perturbed map $f_\ep$ has a smooth invariant measure
in small neighborhood of the point $c$, converging to the delta-measure
at this point as $\ep \to 0$. 
\label{loc-hyp} \end{theorem}

\proof Let $\tau_\alpha:\IR^d\to \IR^d$ be a $d$-dimensional rotation
around the point $c \in \IR^d$ through the angle $\alpha$, such that
$\tau_\alpha E_c^u \subset E_c^s$ and any coordinate of the vector $\alpha$
lies in the region $[-\pi /2, \pi /2]$. For each $\ep >0$ we define
a map $\tau_\alpha^{(\ep)}:\IR^d \to \IR^d$ such that for all 
$x \in B_{\ep /\sqrt N}(c)$, where $N = \dim(E_c^u)$, it coincides with 
$\tau_\alpha$ and in the compliment of this ball it is defined as follows:
$$ \tau_\alpha^{(\ep)} x = \tau_{\ep \alpha /(|x-c| \sqrt N)}x .$$ 
Then 
$$ |x - \tau_\alpha^{(\ep)} x| \le \ep\pi/2 .$$ 
Now we fix some $0 < \beta \ll 1$ and define the following family of 
transition probability densities:
$$ q_\ep(x,y)=\cases{1/|B_{\beta\ep}|, 
                                            &if $y \in B_{\beta\ep}(\tau_\alpha^{(\ep)} x)$; \cr
                                        0, &otherwise. \cr} $$
Evidently, that such perturbations are mixing and due to
the estimate above $q_\ep(x,y)=0$ for $|x-y| > 2\ep$ and $\beta < 0.4$. On 
the other hand just as in Lemma~\ref{local-ex} any trajectory of
the stochastically $\ep$-perturbed map, started from the ball
$A_\ep = \{x \in \IR^d: |x-c| \le \ep /(2\Lambda_u)\}$ remains in it with the
probability $1$. Indeed for 
$$ \beta < 2\Lambda_u(1 + \la_s \Lambda_u)/(1 - \la_s \Lambda_u) $$
we have 
$$ \la_s \Lambda_u (\ep /(2 \Lambda_u) + \beta\ep) + \Lambda_u\beta\ep 
     < \ep /(2\Lambda_u), $$ 
which finishes the proof. Note that the constant $\beta$ here does not
depend on $\ep$. \qed

The generalization of this result for an arbitrary hyperbolic
periodic trajectory is straightforward. 
\subsection{$H_\infty$ versus $H_\infty^\ep$ assumption}
In the case of singular 
perturbations we have to replace the simple hyperbolicity assumption
$H_\infty$ by the stronger one $H_\infty^\ep$, see Theorems
\ref{stab3} and \ref{stab4}. To construct an example satisfying
$H_\infty$ without being stochastically stable,
consider a map $f$ with fixed turning point $c=f(c)$, such that locally 
(in a neighborhood of this point) the map is defined as follows:
\be f(x) :=\cases{\la_1x + (1-\la_1)c, &if $x\le c$ \cr 
                               \la_2x + (1-\la_2)c, &otherwise ,\cr}  \label{fold} \ee
where $|\la_1|>1>|\la_2|>0$ and $|\la_1\la_2|>1$. Both of the maps in 
Figure~\ref{maps2} satisfy these assumptions. We consider the 
following singular perturbations:
\be x \longrightarrow \cases{x + \ep, &with probability $1-q$ \cr 
                                                     x - \ep, &with probability $q$,\cr } 
\label{sing2} \ee
where $0<q<1$ is a parameter. Note that this setting satisfies the
$RW$ assumption.
%
%%%%%%%%%%%%%%%%%%%%%%%%%%%%%%%%%%%%%%%%%%%%%%%%%%%%%%%%%%%%%%%%%%
%% Counter examples
\Bfig(350,150)
    {% picture (a)
     \drawline(0,0)(150,0)(150,150)(0,150)(0,0) 
     \drawline[-50](0,0)(150,150) \drawline[-50](75,0)(75,75)
     \thicklines
     \drawline(0,150)(30,0)(75,75)(120,60)(150,150)
     \put(73,-10){$c$} \put(70,155){{\bf(a)}}
     \put(0,-10){$0$} \put(150,-10){$1$} \put(-10,145){$1$}
     % picture (b)
     \put(200,0){\bpic(150,150){\thinlines
     \drawline(0,0)(150,0)(150,150)(0,150)(0,0) 
     \drawline[-50](0,0)(150,150) \drawline[-50](30,0)(30,30)
                                  \drawline[-50](120,0)(120,120) 
     \put(28,-8){$c$}  \put(110,-8){$1-c$} \put(73,-10){$1/2$}
     \put(70,155){{\bf(b)}}
     \put(0,-10){$0$} \put(150,-10){$1$} \put(-10,145){$1$}
     \thicklines
     \drawline(0,150)(30,30)(75,0) \drawline(75,150)(120,120)(150,0)
     }}
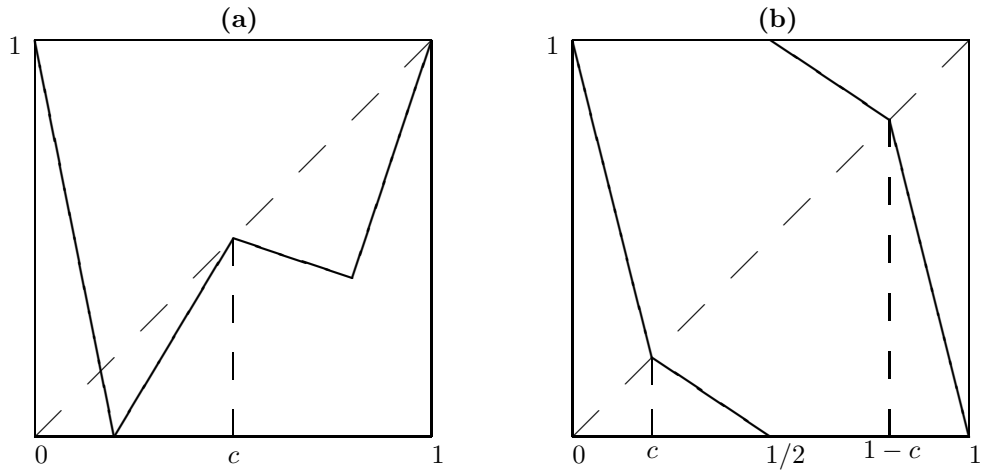
    }{PE maps. (a) nonsimmetric W-map, example of a PE map of type 
      $H_\infty$ \label{maps2}}
%%%%%%%%%%%%%%%%%%%%%%%%%%%%%%%%%%%%%%%%%%%%%%%%%%%%%%%%%%%%%%%%%%
%

Clearly, the map $f(x)+\ep$ has a stable fixed point $c_\ep:=c+\ep/(1-\la_2)$. 
Now let $h \in L^1$ be a nonnegative function, such that $h(c_\ep)>0$. 
Then the transition operator $P$ of the random map satisfies the 
following inequality:
$$ Ph(c_\ep) \ge (1-q) \Lpar \frac1{|\la_1|} h(c + \frac\ep{1-\la_1}) 
                                         + \frac1{|\la_2|} h(c_\ep) \Rpar
     \ge \frac{1-q}{|\la_2|}h(c_\ep) .$$
Therefore for any positive integer $n$ we have
$$ P^nh(c_\ep) \ge \Lpar\frac{1-q}{|\la_2|}\Rpar^n h(c_\ep) .$$
Thus, for $1-q > |\la_2|$ the value of $P^nh$ at the point $c_\ep$ is growing 
exponentially, so an exponential decay assumption in 
$L^\infty$-- or $\BV$--norm that we introduce in Section 
\ref{random-walk-assumption} is violated.

In spite of this, using different methods, we are able to prove the convergence 
to a smooth SBR measure $\mu_\ep$ in this example. Note, that new phenomenon 
is observed here. Indeed, due to the estimates above, the density of the SBR 
measure $\mu_\ep$ is not a function of bounded variation and usual 
Lasota-Yorke type estimates do not work here. The point is that the 
mathematical expectation of the slope of the random map at some points 
is less than $1$ (in our example $E(|f'(c_\ep)|)=|\la_2|<1$), which contradicts 
usual settings (see \cite{{Mo},{Pe}} and references therein). Indeed, if such a 
point is a fixed point for some of the shifted maps (as the point $c_\ep$ for 
the map $f(x)+\ep$ in our example), then this really leads to unusual 
properties of the transition operator of the random map. Investigation of 
ergodic properties of such random maps is out of the scope of the present 
paper and will be published elsewhere.

\section{Proofs of the stability theorems
% \ref{stab2}, \ref{stab3}, and \ref{stab4}}
}
\subsection{The general setting}
\subsubsection{The deterministic part of the dynamics}
Let $X$ be a bounded interval in $\rz$, and denote by $m$ 
Lebesgue measure on $X$. Three function spaces over
$X$ play an important role 
in our studies:
\begin{itemize}
\item
$L^1=L^1_m(X)$, the space of $m$-equivalence classes of complex 
valued integrable functions on $X$, which is endowed with the norm
$\|h\|_1=\int_X |h|\,dm$,
\item
$L^\infty=L^\infty_m(X)$, the space of $m$-equivalence classes of complex 
valued bounded functions on $X$, which is endowed with the norm
$\|h\|_\infty=\mbox{ess sup}_X |h|$, and
\item
$\BV$, the space of $m$-equivalence classes of integrable functions of 
bounded variation on $X$. It is endowed with the norm $\bv{h}=\|h\|_1+\var(h)$
where $\var(h)=\var_{\rz}(h)$ and where for any $Y\subseteq \rz$
\beqas
\var_Y(h)&=&\sup\{\int_X \ph'h\,dm:\ph\in \Test(Y)\}\quad\mbox{and}\\
\Test(Y)&=&\{\ph\in C(\rz):
\|\ph\|_\infty\le 1,\ \ph_{|\rz\setminus Y}=0,
\ph\mbox{ differentiable, }\|\ph'\|_\infty<\infty\}\ .
\eeqas
We remark that the same notion of variation is obtained if one restricts 
$\Test(Y)$ to $C^1$- or even $C^\infty$-functions.  
\end{itemize}         
\begin{bem}
As $X$ is bounded, $\|h\|_\infty\le\frac 12\var(h)$ for all $h\in\BV$,
and if $I\subseteq X$ is an interval, then
$\var(h\cdot 1_I)\le\var(h)$. Indeed, our setting means that $\var(h)$ is the variation of $1_X\cdot h$ over $\rz$.
\end{bem}
We study dynamics on $X$ given by a map $f:X\to X$ composed with a random 
perturbation with transition kernel $Q$. More specifically,
let $f:X\to X$ be piecewise monotone with a finite number of turning points
$\TP:=\{c_1,\ldots,c_r\}$. 
By this we mean that for each maximal component $Z$ of
$X\setminus\TP$ holds
\be\label{basicassumptions}
\pbox
{$f_{|Z}$ is monotone and continuously differentiable
and extends continuously to $\bar Z$,
$\var_Z|f_{|Z}'|<\infty$. In particular, $\Lambda:=\sup_X|f'|<\infty$.}
\ee
These intervals $Z$ form a partition $\Z$ 
of $X$ modulo the finite set $\TP$. Let 
$\Z_n=\{Z_1\cap f^{-1}Z_2\cap\ldots\cap f^{-(n-1)}Z_{n}\neq\emptyset:\,Z_i\in\Z\}$.
Then $\Z_n$ is a partition of $X$ modulo finitely many points, and 
$f^n_{|Z}$ satisfies the basic assumptions (\ref{basicassumptions}) for 
each $Z\in\Z_n$.
\par
Our essential hyperbolicity assumption is:
\be\label{hyperbolicity}
\pbox
{There are 
constants $\lambda>1$, $\eta>0$ and $N\in\nz$ such that
\[
|(f^N_{|Z})'|\ge\lambda\mbox{ for all $Z\in\Z_n$\quad and}\quad
\eta\le |f_{|Z}'|\le\eta^{-1}\mbox{ for all $Z\in\Z$.}
\]
}
\ee
Of course, once we have such a constant $\lambda>1$, we can (and will) 
assume that (for a larger iterate $N$) we even have $\lambda>\hyp$.
\par
Given such an iterate $N$ and a suitable $\beta>0$ that we specify later, 
we refine the partition $\Z$ by adding further points to $\TP$ in such a 
way that 
\be\label{betaassumption}
\var\left(\frac{f_{|Z}'}{\eta}\right),\ \var\left(\frac{\eta^{-1}}{f_{|Z}'}
\right)
\le
\beta \ .
\ee
We further modify the set $\TP$ by doubling all its elements together with 
their preimages and extend $f$ to the enlarged space by one-sided 
continuity. We call the resulting space, the extended
map and the collection of doubled turning points again $X$, $f$, and $\TP$
respectively. (Strictly speaking the new space $X$ is a linearly ordered, 
order complete space. Its order topology will be quite different from the 
topology on the real line, usually it will be totally disconnected.)
\par
If after these modifications 
there are $c_i,c_j\in\TP$ such that $f^kc_i=c_j$ for some $k>0$, we 
also add all points $f^lc_i$, $l=1,\ldots,k-1$, to $\TP$. In this way we 
can make sure that 
\be\label{periodicity}
\mbox{for each $c_i\in\TP$ either $fc_i\in\TP$ or $f^kc_i\not\in\TP$ for 
all $k\ge 1$.}
\ee
We denote the enlarged set $\TP$ once more by $\TP=\{c_1,\ldots,c_r\}$.
For $c_j\in\TP$ we mean by $c_{\hat j}\in\TP$ its doubled copy. 
\par
Let $J_j:=[c_j,d_j]\subset\tilde J_j:=[c_j,\tilde d_j]$, $j=1,\ldots,r$ be
two families of one-sided interval neighbourhoods 
of the turning points. Here as in the sequel 
we write $[u,v]=\{x\in X: u\le x\le v\mbox{ or }v\le x\le u\}$, in particular 
$[u,v]=[v,u]$. 
The 
$\tilde J_j$ are chosen such that any two of them are disjoint. 
(Observe that this 
holds also for $\tilde J_j$ and $\tilde J_{\hat j}$!) 
We write $Y=\cup_j J_j$ and $\tilde Y=\cup_j\tilde J_j$ and assume 
that for each $j=1,\ldots,r$ holds
\be\label{Jcycle}
\pbox
{\begin{itemize}
\item
either $f(J_j)\cap \tilde Y=\emptyset$
\item
or $f(J_j)\supset \tilde J_i$ 
for some $i$ in such a way that $f(c_j)=c_i$ 
but $f(d_j)\not\in \tilde J_i$.
\end{itemize}
}
\ee
Observe that this includes a kind of topological expansion assumption for 
neighbourhoods of periodic turning points. If $f\in H_\infty$, 
then this assumption can always be satisfied.
\par
On the level of ``mass transport'' the dynamics of $f$ are described by 
the {\em Perron-Frobenius operator} $P_f:L^1\to L^1$,
\[
P_fh=\sum_{Z\in\Z} \frac{h}{|f'|}\circ f_{|Z}^{-1}\cdot 1_{fZ}\ .
\]
We note that $\int P_fh\,dm=\int h\,dm$ for all $h\in L^1$.
Later we shall see that $P_f$ is also a bounded linear operator on $\BV$.
\subsubsection{The stochastic part of the dynamics}\label{stochastic-part}
Perturbations are described by (sub)-Markovian operators $Q:L^1\to L^1$. More 
precisely we assume
\begin{enumerate}
\item
$Q$ is a linear, positive operator and $\|Q\|_1\le 1$.
\item
$Q^*:L^\infty\to L^\infty$ denotes the dual operator to $Q$.
\end{enumerate}         
We remark that $Q$ can always be thought of as represented by a 
(sub)-Markov transition kernel, i.e. 
\[
Q^*f(x)=\int f(y)\,Q(x,dy)\quad\mbox{and}\quad 
Qh(y)=\left(\frac d{dm}\int h(x)\,Q(x,.)m(dx)\right)(y)\ ,
\]
see {\em e.g.} \cite[\S 3.1]{Krengel}.
If $Q(x,.)\ll m$ for each $x$ (what our assumptions do not necessarily 
imply), we denote $q(x,y)=\frac d{dm}Q(x,.)(y)$.
\par
The following regularity assumption on $Q$ plays an essential role:
There is a constant $C>0$ such that
\beqa\label{QLYassumption}
\var(Qh)
&\le&
\var(h)+C\cdot\|h\|_1\quad \mbox{and}\\
\var_{J_k\cup J_{\hat k}}(Qh)
&\le&
\var_{\tilde J_k\cup \tilde J_{\hat k}}(h)
+C\cdot\|h\cdot 1_{\tilde J_k\cup\tilde J_{\hat k}}\|_1 \nonumber 
\eeqa
for each $h\in\BV$
and $k\in\{1,\ldots,r\}$. As we shall see later, 
this assumption is satisfied in many cases of interest, including 
absolutely continuous, random walk-like and also deterministic
perturbations.
\par
Finally we define the {\em spread} of $Q$ as
\[
\spread(Q)=\inf\{\delta>0:\ Q(x,\{y:|y-x|>\delta\})=0\ \forall\,x\}\ .
\]
The $\spread$ is the size of the largest jump that can be caused by the 
perturbation. We assume that
\be\label{spreadassumption}
\spread(Q)<\min\{|\tilde d_j-d_j|:j=1,\ldots,r\}\ ,
\ee
i.e. random jumps from $X\setminus\tilde Y$ to $Y$ are excluded.
\subsubsection{The decomposition}         
In the sequel we study the behaviour of $f$ on 
$Y$ and on $X\setminus Y$ separately under the assumption that transitions 
from $X\setminus Y$ to $Y$ are restricted in the following sense:
Define $\tilde P_1,\tilde P_2:BV\to BV$ by
\be\label{decomposition}
\tilde P_1 h=QP_f(h\cdot 1_{X\setminus Y}),\quad
\tilde P_2 h=QP_f(h\cdot 1_Y)\ .
\ee
A straightforward calculation shows
\begin{proposition}\label{P1-P2-prop}
Suppose that there are constants $C_1,C_2>0$ and $\alpha\in(0,1)$ such 
that
\be\label{LY-Ungleichung}
\var(\tilde P_j^kh)\le C_1\alpha^k\var(h)+C_2\|h\|_1\quad\mbox{ for all 
}k\in\nz
\ee
and that there is some $N\in\nz$ such that 
\be\label{transitions}
\tilde P_2\tilde P_1^k\tilde P_2=0\quad\mbox{ for all $k=1,\ldots,N$.}
\ee
Then
\be\label{LY-global}
\var((QP_f)^N h)
\le
\left(\frac{N(N+1)}{2}C_1^3+C_1\right)\alpha^N\cdot\var(h)
+(1+C_1+C_1^2)C_2\frac{(N+1)^2}2\cdot\|h\|_1\ .
\ee
\end{proposition}
In order to guarantee (\ref{transitions}) for a fixed $N$ the intervals 
$J_j$ and the spread of $Q$ must be taken sufficiently small:
\begin{lemma}\label{N-lemma}
Given $f$, $N$ and the set $\TP$, there is $\delta>0$ such that 
(\ref{transitions}) is satisfied if $\spread(Q)<\delta$ and
$|J_j|<\delta$ for all $j=1,\ldots,r$.
\end{lemma}
\proof
Consider a trajectory of the perturbed system 
$x_0,x_1,\ldots,x_k$ that starts at $x_0\in J_i\subseteq Y$ 
and suppose that it 
ends at $x_k\in J_j\subseteq Y$
with $x_l\not\in X\setminus Y$ for $l=1,\ldots,k-1$. 
If $\delta$ in the assumptions 
of the lemma is small, the trajectory stays close to a 
sequence $c_i=z_0,z_1,\ldots,z_k$ satisfying 
for each $l=1,\ldots,k$ either $z_{l}=fz_{l-1}$ or
$z_{l}\in\{c_m,c_{\hat m}\}$ if $c_m=fz_{l-1}$. 
In particular, making $\delta$ small we can force $x_k$ to be as close as we 
like to the point $z_k\in\cup_{l=1}^kf^l(\TP)$, and as $x_k\in J_i$, $z_k$ 
is at the same time close to $c_j$. Because of (\ref{periodicity})
this is possible only if $z_0,\ldots,z_k\in\TP$.
But this is incompatible with the assumptions (\ref{Jcycle})
and (\ref{spreadassumption}).
\qed
\subsubsection{The scheme of the proofs}\label{scheme-of-proofs}
Given not only one perturbation operator but a whole family 
$Q_\ep$ ($\ep>0$), we will have to
show that there are decompositions $X=Y\cup (X\setminus Y)$ of $X$
depending on $\ep$, 
such that the corresponding operators 
$\tilde P_{1,\ep}$ and $\tilde P_{2,\ep}$
satisfy (\ref{LY-Ungleichung}) uniformly in $\ep$ ({\em i.e.} with constants
$C_1,C_2,\alpha$ not depending on $\ep$). Then there is $N>0$ such that $\left(\frac{N(N+1)}2 C_1^3+C_1\right)\,\alpha^N<\frac 12$, 
and according to Lemma \ref{N-lemma}
and Proposition \ref{P1-P2-prop} there are constants $\delta,C_3>0$ such that
\be\label{LY-12}
\var((Q_\ep P_f)^N h)
\le
\frac 12\cdot\var(h)+ C_3\cdot\|h\|_1\quad\mbox{for all $h\in\BV$.}
\ee
As usual this implies that each $Q_\ep P_f$ has an invariant density $h_\ep$
and $\var(h_\ep)\le 2C_3$ for all $\ep$. In particular, the family
$(h_\ep)_{\ep>\delta}$ is relatively compact in $L^1$. Consider any limit
$h=\lim_{i\to\infty}h_{\ep_i}$ with $\ep_i\to 0$.
The invariance $Q_\ep P_f h_\ep=h_\ep$ follows immediately
from
\[
\|P_f h-h\|_1
\le
\|P_f h-E_\ep P_f h\|_1+\|Q_\ep P_f h-Q_\ep P_f h_\ep\|_1
+\|h_\ep-h\|_1
\]
and the assumption that $\|Q_\ep P_f h-P_f h\|_1\to 0$ as $\ep \to 0$.
We specialize this to
\begin{lemma}\label{general-lemma}
If $P_f$ has a unique invariant density $h$, then, under the above assumptions
on the family $Q_\ep$, the invariant densities $h_\ep$ of $Q_\ep P_f$
converge to $h$ in $L^1$ as $\ep\to 0$.
\end{lemma}
This lemma reduces the proofs of Theorems
\ref{stab2}, \ref{stab3}, \ref{stab4} and \ref{ulam}
to proving (\ref{LY-Ungleichung}) uniformly in $\ep$ under the various assumptions of these four theorems.
Much more detailed informations about the convergence $h_\ep\to h$ and about the
ergodic properties of the perturbed systems $f_\ep$ can be gained 
by exploiting more carefully the spectral theoretic consequences
of (\ref{LY-12}) in the light of the Ionescu-Tulcea/Marinescu theorem, see {\em e.g.} \cite{Ke}.
\subsection{Estimates far from turning points}
In this section we prove (\ref{LY-Ungleichung}) for $\tilde P_1$, i.e.\ 
for branches with orbits far from turning points. As an immediate corollary we
obtain Theorem \ref{stab2}.
\subsubsection{Interchanging the map and the perturbation}
Fix a sequence of monotone branches $f_i=f_{|Z_i\setminus Y}$, 
$i=1,\ldots,N$ of $f$ and 
study the operator
$QP_{f_{1}}QP_{f_{2}}\dots QP_{f_N}$.
For each of the $i=1,\ldots,N$ we fix a monotone and 
continuously differentiable extension 
$\tilde f_i:\rz\to \rz$ of $f_i$.
Given $\beta>0$ as in (\ref{betaassumption}) one can always find such an 
extension with the additional property that
there is $y_0\in Z_i$ such that
\be\label{varbeta}
\var\left(\frac{\ff'}{f'(y_0)}\right),\ \var\left(\frac{f'(y_0)}{\ff'}\right)
\le\beta\ .
\ee
\begin{lemma}
Let $Q$ be a (sub)Markov operator satisfying 
\[
\mbox{$\var(Qh)\le\var(h)+C\cdot\|h\|_1$ for all $h\in\BV$.}
\]
For a fixed branch $f_i$ of $f$ as above
define $\tilde Q:=P_{\ff_i^{-1}}QP_{f_i}$. Then 
$P_{\ff_i}\tilde Q=QP_{f_i}$ and
\[
\var(\tilde Qh)
\le
(1+ \frac 32 \beta)^2\cdot\var(h\cdot 1_I)
+C(1+ \frac 32 \beta)\|f'\|_\infty\cdot\|h\cdot 1_I\|_1
\]
for all $h\in\BV$.
\end{lemma}
\proof
For notational convenience we write $f$ and $\ff$ instead of $f_i$ and 
$\ff_i$, and we denote $I= Z_i\setminus Y$.
Let $h\in\BV$, $y\in X$ and $u(y):=\frac{\ff'(y)}{f'(y_0)}$. Then
$\var(u)\le\beta$, $\|u\|_\infty\le 1+\beta$ and
\[
\tilde Qh(y)
=
\frac{1}{|(\ff^{-1})'(\ff y)|}\cdot QP_fh(\ff y)
=
u(y)\cdot QP_f(f'(y_0)\cdot h)(\ff y)
\]
because $QP_f$ is a linear operator.
Therefore
\beqas
&&\var(\tilde Qh)\\
&\le&
\var(u)\cdot\|QP_f(f'(y_0)\cdot h)\|_\infty
+\|u\|_\infty\cdot\var(QP_f(f'(y_0)\cdot h))\\
&\le&
\beta\cdot\frac 12\var(QP_f(f'(y_0)\cdot h))
+(1+\beta) \cdot \var(QP_f(f'(y_0)\cdot h))\\
&\le&
(1+ \frac 32 \beta)\cdot\left(\var(P_f(f'(y_0)\cdot h))
+C\cdot\|P_f(f'(y_0)\cdot h)\|_1\right)\\
&\le&
(1+ \frac 32 \beta)\cdot\left(\var((\frac{f'(y_0)}{\ff'}\cdot h)\circ f^{-1}
\cdot 1_{fI})
+C\cdot\|f'(y_0)\cdot h\cdot 1_I\|_1\right)\\
&\le&
(1+ \frac 32 \beta)\cdot\left(\var(\frac{f'(y_0)}{\ff'}\cdot h\cdot 1_I)
+C\cdot\|f'\|_\infty\cdot \|h\cdot 1_I\|_1\right)\\
&\le&
(1+ \frac 32 \beta)\cdot\left(\var(\frac{f'(y_0)}{\ff'})
\|h\cdot 1_I\|_\infty
+\|\frac{f'(y_0)}{\ff'}\|_\infty\var(h\cdot 1_I)\right.\\
&&\hspace*{4cm}
\left.\frac{}{}+C\cdot\|f'\|_\infty\cdot \|h\cdot 1_I\|_1\right)\\
&\le&
(1+ \frac 32 \beta)\cdot\left(\beta\cdot\frac 12\var(h\cdot 1_I)
+(1+\beta)\cdot\var(h\cdot 1_I)
+C\cdot\|f'\|_\infty\cdot \|h\cdot 1_I\|_1\right)\\
&\le&
(1+ \frac 32 \beta)^2\var(h\cdot 1_I)
+(1+ \frac 32 \beta)C\|f'\|_\infty\cdot\|h\cdot 1_I\|_1
\eeqas
\qed
Next we apply this lemma to the sequence 
$f_1,\ldots,f_{N}$ of branches of $f$.
\begin{lemma}
Let $Q_1,\ldots,Q_{N}$ be 
(sub)Markov operators with 
\[
\var(Q_j h)\le\var(h)+C_j\|h\|_1
\]
for some constants 
$C_j>0$ ($j=1,\ldots,N$). 
Then there is for each $k=1,\ldots,N$ a (sub)Markov operator $\tilde Q_k$ 
such that
\[
Q_1P_{f_1}Q_2P_{f_{2}}\dots Q_kP_{f_k}
=
P_{\ff_1}P_{\ff_{2}}\dots P_{\ff_k}\tilde Q_k
\]
and for each $h\in\BV$
\[
\var(\tilde Q_k h)\le (1+ \frac 32 \beta)^{2k}\cdot\var(h)+\tilde C_k\cdot\|h\|_1
\]
where 
\[
\tilde C_k=\sum_{j=1}^k (1+ \frac 32 \beta)^{2(j-1)}C_j
\prod_{i=j}^k[(1+ \frac 32 \beta)\|f_j'\|_\infty]\ .
\]
\end{lemma}
\proof
The proof is by induction. For $k=1$ this is just the statement of the 
previous lemma. Suppose the lemma is true for $k=n$. Then
\[
Q_1P_{f_{1}}Q_2P_{f_{2}}\dots Q_nP_{f_n}Q_{n+1}P_{f_{n+1}}
=
P_{\ff_{1}}P_{\ff_{2}}\dots P_{\ff_n}\tilde Q_nQ_{n+1}P_{f_{n+1}}
\]
by inductive hypothesis. The previous lemma applied to $P_f=P_{f_{n+1}}$
and $Q=\tilde Q_nQ_{n+1}$ 
yields the existence of $\tilde Q_{n+1}$ such that
$\tilde Q_nQ_{n+1}P_{f_{n+1}}=P_{\ff_{n+1}}\tilde Q_{n+1}$.
Since 
\beqas
\var(\tilde Q_n Q_{n+1}h)
&\le&
(1+ \frac 32 \beta)^{2n}\cdot\var(Q_{n+1}h)+\tilde C_n\cdot\|Q_{n+1}h\|_1\\
&\le&
(1+ \frac 32 \beta)^{2n}\cdot\var(h)
+((1+ \frac 32 \beta)^{2n} C_{n+1}+\tilde C_{n})\cdot\|h\|_1\ ,
\eeqas
we have
\[
\var(\tilde Q_{n+1}h)
\le
(1+ \frac 32 \beta)^{2(n+1)}\var(h)+(1+ \frac 32 \beta)\|f_{n+1}'\|_\infty(\tilde 
C_n+(1+ \frac 32 \beta)^{2n}C_{n+1})\cdot\|h\|_1\ .
\]
Inserting the formula for $\tilde C_n$ finishes the proof.
\qed
Now we combine this lemma with the hyperbolicity assumption
(\ref{hyperbolicity}) for $f$:
\begin{lemma}\label{Lemma4}
Suppose that in the situation of the previous lemma
there are 
constants $\lambda>1$, $\eta>0$ such that
\[
|(f^N_{|Z})'|\ge\lambda\mbox{ for all $Z\in\Z_n$\quad and}\quad
|f_{|Z}'|\ge\eta\mbox{ for all $Z\in\Z$.}
\]
(This is assumption (\ref{hyperbolicity}).)
If $f_1\circ\dots\circ f_N$ is not the empty map (i.e.\ if 
$f_1\circ\dots\circ f_N$ is part of a branch of $f^N$),
then
\[
\var(Q_1P_{f_1}Q_2P_{f_{2}}\dots Q_NP_{f_N}h)
\le
\left((1+ \frac 32 \beta)^{2N}\cdot\lambda^{-1}+\frac 12 N\beta\eta^{-N}\right)
\cdot\var(h)
+\lambda^{-1}\tilde C_N\cdot\|h\|_1\ .
\]
for each $h\in\BV$.
\end{lemma}
\proof
Because of the variation condition (\ref{varbeta})
\[
\var(\frac 1{\ff_j'})\le\beta\cdot\|\frac 1{\ff_j'}\|_\infty
\le\frac \beta\eta\ .
\]
Therefore a simple induction yields
\[
\var\left(\frac 1{|(\ff_1\circ\ldots\circ\ff_N)'|}\right)
\le N\beta\eta^{-N}\ .
\]
Hence
\beqas
&&\var(Q_1P_{f_1}Q_2P_{f_{2}}\dots Q_NP_{f_N}h)\\
&=&
\var(P_{\ff_1}P_{\ff_{2}}\dots P_{\ff_N}\tilde Q_Nh)\\
&=&
\var\left(\frac {\tilde Q_N h}{|(\ff_1\circ\ldots\circ\ff_N)'|}
\circ(\ff_1\circ\ldots\circ\ff_N)^{-1}\right)\\
&\le&
\var(\tilde Q_N h)\cdot\lambda^{-1}
+\underbrace{\|\tilde Q_N h\|_\infty}_{\le\frac 12\var(\tilde Q_N h)}
\cdot N\beta\eta^{-N}\\
&\le&
\left((1+ \frac 32 \beta)^{2N}\cdot\lambda^{-1}+\frac 12 N\beta\eta^{-N}\right)
\cdot\var(h)
+\lambda^{-1}\tilde C_N\cdot\|h\|_1\ .
\eeqas
\qed
\begin{lemma}
Let
\[
Z=Z_N\cap f_N^{-1}Z_{N-1}\cap(f_{N-1}\circ f_N)^{-1}Z_{N-2}\ldots\cap
(f_2\circ\dots\circ f_N)^{-1}Z_1\ ,
\]
and as before $f_i=f_{|Z_i\setminus Y}$.
(Note that $Z\in\Z_N$ if $Z\neq\emptyset$.)
Recall that $\Lambda=\sup|f'|<\infty$. If 
\[
\spread(Q)<\frac{\Lambda-1}{\Lambda^{N}}\cdot\min\{|J_i|:i=1,\ldots,r\}\ ,
\]
then
\[
QP_{f_1}\dots QP_{f_N}(h)=QP_{f_1}\dots QP_{f_N}(h\cdot 1_Z)
\]
for each $h\in\BV$.
\end{lemma}
\proof
Let $Z'\in\Z_N$, $Z'\neq Z$. We prove that
\be\label{pathproperty}
QP_{f_1}\dots QP_{f_N}(h \cdot 1_{Z'})=0\ .
\ee
Suppose this is not true. 
Then there is at least one $\delta$-pseudoorbit 
$x_N,x_{N-1},\ldots,x_1,x_0$ with
\begin{itemize}
\item
$\delta=\spread(Q)$,
\item
$x_N\in Z'$, in particular $f^{N-j}(x_N)\in Z_j'$ ($j=N,\ldots,1$),
\item
$x_j\in Z_j\setminus Y$ ($j=N,\ldots,1$), in particular $Z_N'=Z_N$, and
\item
$|x_{j-1}-f_j(x_j)|<\delta$ ($j=N,\ldots,1$). 
\end{itemize}         
It follows that
\[
|x_{j}-f^{N-j}(x_N)|<\sum_{i=0}^{N-j-1}\delta\Lambda^i
<\delta\frac{\Lambda^N}{\Lambda-1}<\min\{|J_i|:i=1,\ldots,r\}
\]
for $j=N,\ldots,1$ such that $x_j\in Z_j'\cup Y$.
But this is possible only if $Z_j=Z_j'$ ($j=N,\ldots,1$), i.e.
if $Z=Z'$, a contradiction.
\qed
\begin{proposition}
If the assumptions
(\ref{basicassumptions}),
(\ref{hyperbolicity}),
(\ref{betaassumption}),
(\ref{QLYassumption}), and
(\ref{spreadassumption}) are satisfied, then there are 
$C_1,C_2>0$, $\alpha\in(0,1)$ and $\delta>0$ 
such that for all $h\in\BV$ and all $k\in\nz$
\[
\var(\tilde P_1^kh)\le C_1\alpha^k\cdot\var(h)+C_2\cdot\|h\|_1
\]
if $\spread(Q)<\delta$ and
$|J_j|<\delta$ for all $j=1,\ldots,r$.
In fact, $C_1=\left(\frac 2\eta\right)^N$, 
$\alpha=\left(\frac 34\right)^{1/N}$, $C_2=C_2(N,C,f)$.
\end{proposition}
\proof
Let $M:\BV\to\BV$ denote multiplication by $1_{X\setminus Y}$.
In view of the previous lemma we can expand $\tilde P_1^N(h)$ as 
\beqas
\tilde P_1^N(h)
&=&
(QP_fM)^N(h)
=\sum_{Z_1,\ldots,Z_N\in\Z} QP_{f_{Z_1}}M\dots QP_{f_{Z_N}}M(h)\\
&=&\sum_{Z\in\Z_N} 
Q(P_{f_1}Q)\dots (P_{f_{N-1}}Q)P_{f_{N}}(h\cdot 1_Z)\ .
\eeqas
Hence, by Lemma \ref{Lemma4}
\beqas
\var(\tilde P_1^N(h))
&\le&
\sum_{Z\in\Z_N}\left[
\left((1+ \frac 32 \beta)^{2N}\cdot\lambda^{-1}
+\frac 12 N\beta\eta^{-N}\right)
\cdot\var(h\cdot 1_Z)
+\lambda^{-1}\tilde C_N\cdot\|h\cdot 1_Z\|_1\right]\\
&\le&
\sum_{Z\in\Z_N}\left[
\left(2(1+ \frac 32 \beta)^{2N}\cdot\lambda^{-1}+N\beta\eta^{-N}\right)
\cdot\var_Z(h)\right.\\
&&\hspace{1.5cm}
\left.
+\max\{|Z|^{-1}:Z\in\Z_N\}\lambda^{-1}\tilde C_N\cdot\|h\cdot 1_Z\|_1\right]\\
&\le&
\left(2(1+ \frac 32 \beta)^{2N}\cdot\lambda^{-1}+N\beta\eta^{-N}\right)
\cdot\var(h)
+\max\{|Z|^{-1}:Z\in\Z_N\}\lambda^{-1}\tilde C_N\cdot\|h\|_1\ .
\eeqas
Now choose $\beta$ so small that 
$2(1+ \frac 32 \beta)^{2N}\cdot\lambda^{-1}+N\beta\eta^{-N}<(3\lambda^{-1})^{\frac 1N}$
and observe that $\lambda>\hyp$. This proves the proposition for $k=N$ 
with $C_1=1$, $\alpha=\left(\frac 34\right)^{1/N}$ and some 
$C_2=C_2(N,C,f)$. Iterated application of this inequality extends it to 
integer multiples $k$ of $N$ with a new 
$C_2=\frac 1{1-\alpha}C_{2,{\rm old}}$. The extension to general $k$ 
follows from the elementary inequality
\[
\var(\tilde P_1h)
\le
\frac 2\eta\var(h\cdot 1_{X\setminus 
Y})+(C+\const)\cdot\|h\|_1
\le
\frac 4\eta\var(h)+(C+\const)\cdot\|h\|_1
\]
with a constant $\const$ depending only on $f$. 
\qed
\subsubsection{Proof of Theorem 1.2}
If $f$ has no periodic turning points, then $\tilde P_2^k=0$ for some 
$k\in\nz$ that depends only on $f$, 
such that (\ref{LY-Ungleichung}) is trivially satisfied for 
$\tilde P_2$ with constants $C_1,C_2,\alpha$ depending only on $f$ and $Q$.
In view of Lemma \ref{general-lemma} this implies Theorem \ref{stab2}.
\subsection{Estimates close to turning points}\label{close}
In this section we prove (\ref{LY-Ungleichung}) for $\tilde P_2$, i.e.\ 
for branches with orbits close to turning points.
\subsubsection{The random walk assumption}\label{random-walk-assumption}
In order to prevent perturbed trajectories from being trapped
in small neighbourhoods of periodic turning points (this was the case in the
example from Lemma \ref{local-ex}), we introduced the {\em random walk} assumption $RW$ in (\ref{RW-assumption}). Here we relate it to a more
technical {\em exponential decay} assumption. 
\par
Consider $P_f$ and $Q_\ep$ ($\ep>0$) together with a decomposition 
$Q_\ep P_f=\tilde P_1+\tilde P_2=\tilde P_{1,\ep}+\tilde P_{2,\ep}$ as in Section \ref{scheme-of-proofs}.
Let
$*\in\{L^1,L^\infty,\BV\}$. We say that
\be\label{ED-assumption}
\pbox{
$\tilde P_2$ satisfies $\ED_*$, if there are constants $C>0$ and 
$\omega\in(0,1)$
such that
$\|\tilde P_{2,\ep}^n\|_*\le C \omega^n$ for all $n>0$.
}
\ee
In order to derive $\ED_{L^1}$ from $RW$, the intervals $J_k$,
whose union makes up $Y$, must be chosen more carefully than in (\ref{Jcycle}) 
and  (\ref{spreadassumption}) where we only took care 
that $f(J_j)\supset \tilde J_i$ for some $i$ if 
$f(J_j)\cap\tilde Y\neq\emptyset$ and that random jumps from 
$X\setminus\tilde Y$ to $Y$ are excluded. In particular, 
$\spread(Q)\le |fJ_j|-|J_j|$. Now we require that this inequality is in fact nearly an equality. More exactly: Let $Y$ be the smallest domain such that
\[
f_\ep(B_\ep(Y)\backslash Y) \cap Y = \emptyset.
\]
\begin{proposition}\label{RW-ED-lemma}
If $Q_\ep$ satisfies $RW$ and if $f\in H_\infty$, then $\tilde P_2$ satisfies
$\ED_{L^1}$.
\end{proposition}
\par
Consider a periodic turning point $c$ with a period $N$. We assume first
that the map $f$ is locally linear. Let us denote 
$$\la=\min\{\prod_{k=1}^{2N}|f'(f^kx)|:  \;  x,f^Nx \in B_\ep(c)\}.$$

\begin{lemma} Let $f \in H_\infty, \; f^N(c)=c, \; Q_\ep \in RW$. 
Then $ \CP\{f_\ep^{2nN}(x) \notin Y\}\ge \delta^{2nN}$ for each $x\in Y$ and
$n=-\ln(\theta(\la-1))/\ln\theta$.
\label{rw-l} \end{lemma}

\proof The domain $Y$ consists of intervals around the points of the trajectory 
of the periodic turning point $c$. Let us denote the interval around the 
point $c$ by $J$. Then the distance from its boundary points to the point $c$ 
is not more, than $\ep/(\la-1)$. 

Note, that $\la>1$. Indeed, after $N$ iterations any point from the 
neighborhood of the point $c$ either will return to the neighborhood, or will 
leave the interval $J$ (if the map is discontinuous at the point $c$). Therefore 
if $\la\le1$, then  the $H_\infty$-assumption does not hold.

Let us fix a point $x \in J$. Then, applying $H_\infty$ and $RW$ assumptions, 
we have
$$ \CP\{\dist(f_\ep^{2nN}x, c) > \theta \ep \la^n \} \ge \delta^{2nN} $$
for any integer $n$ and for any point $x \in J$. Therefore, if 
$$ \theta \ep \la^n > \frac{\ep}{\la-1}, $$
then any point will leave the interval $J$ in $2nN$ iterations  with some 
positive probability, which will not depend on $\ep$. The value of $n$ 
in the statement of lemma satisfies this inequality. \qed
\begin{corollary}
For small $\ep$ the conclusion of the lemma remains true for piecewise $C^2$--maps, if the
constant $\lambda$ is replaced by a slightly smaller one.
\end{corollary}
Now to finish the proof of the proposition, we need 
the following simple statement.

\begin{lemma} \label{integral} Let $Q$ be any transition kernel satisfying $Q(x,X \backslash Y) \ge \delta$ for 
any point $x \in Y$. Then for any nonnegative function $h \in L^1$ with 
the support in the set $Y$ we have:
$$ \int_Y Qh(x) \, dx \le (1-\delta) \int_Y h(x) \, dx .$$
\end{lemma}

\proof Recall that for any measurable set $A$ we have 
$$ \int_A Qh(x) \, dx = \int Q(x,A) h(x) \, dx. $$
Thus 
$$ \int_Y Qh(x) \, dx = \int Q(x,Y) h(x) \, dx 
          \le (1-\delta) \int_Y h(x) \, dx .$$
\qed
\subsubsection{Smooth perturbations}
We start with a special case of absolutely continuous perturbations for which 
the calculations are particularly easy.
\begin{proposition}
Suppose that $Q$ is an integral operator with kernel $q(x,y)$ satisfying
\[
\var(y\mapsto q(x,y))\le \frac M{|Y|}\quad\mbox{for all $x$}
\]
and such that $\tilde P_2$ satisfies $\ED_{L^1}$.
Then 
\[
\var(\tilde P_2^nh)\le M C \omega^{n-1}\cdot\left(\var(h)+\frac 1{|X|}\|h\|_1\right)
\]
for $h\in\BV$.
\end{proposition}
\proof
Let $h\in\BV$ and denote $\tilde h:=P(h1_Y)$. Then $\tilde P_2 h=Q\tilde h$ and
\beqas
\var(\tilde P_2 h)
&=&
\var\left(y\mapsto\int\tilde h(x)\,q(x,y)\,dx\right)\\
&=&
\sup_{\phi\in\Test(X)}\int\int\tilde h(x)\,\phi'(y)\,q(x,y)\,dx\,dy\\
&\le&
\int|\tilde h(x)|\cdot\left|\sup_{\phi\in\Test(X)}\int\phi'(y)\,q(x,y)\,dy\right|\,dx\\
&\le&
\int|\tilde h(x)|\cdot\var(y\mapsto q(x,y))\,dx\\
&\le&
\frac M{|Y|}\int|P(h1_Y)(x)|\,dx\\
&\le&
\frac M{|Y|}\int_Y|h(x)|\,dx\ .
\eeqas
As $\tilde P_2 h=QP_f(h 1_Y)=\tilde P_2(h1_Y)$, it follows that
\beqas
\var(\tilde P_2^nh)
&\le&
\frac M{|Y|}\int|\tilde P_2^{n-1}h(x)|\,dx\\
&\le&
\frac M{|Y|}\int\tilde P_2^{n-1}|h(x)1_Y(x)|\,dx\\
&\le&
\frac M{|Y|}\cdot\|\tilde P_2^{n-1}\|_1\cdot\int_Y|h(x)|\,dx\\
&\le&
MC\omega^{n-1}\cdot\|h\|_\infty\\
&\le&
MC\omega^{n-1}\cdot\left(\var(h)+\frac 1{|X|}\|h\|_1\right)
\eeqas
\qed
Theorem \ref{stab3} follows from this proposition and Proposition \ref{RW-ED-lemma}
via Lemma \ref{general-lemma}.
\subsubsection{General perturbations}
We now turn to more general perturbations, which are much more delicate.
Our first goal 
is to decompose the operator $\tilde P_2$ as $\tilde P_2=L+\tilde P_2K$, 
where
\[
\var_Y(L^Nh)\le\alpha\cdot\var_Y(h)\quad\mbox{for 
}h\in\BV(Y)
\]
with some suitable constant $0<\alpha<1$ and where
$K$ is a finite rank operator such that also $\tilde P_2K$ is of finite rank. 
Our choice of $K$ is
\[
K h=\sum_{j=1}^r h(c_j)1_{J_j}\ .
\]
As our $\BV$-functions are in fact $L^1$-equivalence classes, one should 
interpret $h(c_j)$ as 
$
\frac 12(\mbox{ess-lim}_{x\nearrow c_j}h(x)
+\mbox{ess-lim}_{x\searrow c_j}h(x))
$.
For later use we note that
\[
L(h\cdot 1_{X\setminus Y})
=
\tilde P_2(h\cdot 1_{X\setminus Y})
-\sum_{j=1}^r h(c_j)\cdot\tilde P_2 1_{J_j}
=
0\ .
\]
\par
For the formulation of the next lemma we introduce the following notation:
Let $A,B\subseteq\{1,\ldots,r\}$. Then
\[
A \to B\quad\mbox{ if }\quad
\exists i\in A, j\in B:\,f(c_i)=c_j\ .
\]
For $i_0\in\{1,\ldots,r\}$ let 
\beqas
\Lambda_N(i_0)
&:=&
\min\{\prod_{l=0}^{N-1}|f'(c_{i_l})|:\,
i_0\to\{i_1,\hat i_1\},\,
\{i_l,\hat i_l\}\to \{i_{l+1},\hat i_{l+1}\}\ (l=1,\ldots,N-1)\}\\
\Lambda_N
&:=&
\min\left\{\Lambda_N(i_0):\,i_0\in\{1,\ldots,r\}\right\}
\eeqas
and observe that
\be\label{Lambda-product}
\Lambda_{N+1}(i_0)
=
|f'(c_{i_0})|\cdot
\min\left\{\Lambda_N(i_1):\,i_0\to\{i_1,\hat i_1\}\right\}
\ee
\begin{lemma}
Fix $N\in\nz$ and $\kappa>1$. There is $\delta>0$ depending only on $f$ and $C$
such that for a refined partition
$\Z$ with $\diam(\Z)<\delta$ holds:
\[
\var_Y(L^Nh)
\le
\sum_{j=1}^r\var_Y(L^N(h\cdot 1_{J_j}))
\le
\kappa\Lambda_N^{-1}\cdot\var_Y(h)\ .
\]
\end{lemma}
\proof
For $j,k\in\{1,\ldots,r\}$
\beqas
\var_{J_k\cup J_{\hat k}}(L(h1_{J_j}))
& = &
\var_{J_k\cup J_{\hat k}}(QP_f((h-h(c_j))\cdot 1_{J_j}))\\
&\le&
\var_{\tilde J_k\cup \tilde J_{\hat k}}
(P_f[(h-h(c_j))\cdot 1_{J_j}])+C\cdot
\|P_f[(h-h(c_j))\cdot 1_{J_j}]\cdot 1_{\tilde J_k\cup\tilde J_{\hat k}}\|_1\\
&=&
\var_{\tilde J_k\cup \tilde J_{\hat k}}(\frac{h-h(c_j)}{|f'|}
\circ f_j^{-1}\cdot 1_{fJ_j})+C\cdot
\|(h-h(c_j))\cdot 1_{J_j}\cdot 1_{\tilde J_k\cup\tilde J_{\hat k}}\|_1 
\eeqas
by(\ref{QLYassumption}) and the fact that $P_f$ is a $L^1$-contraction.
By (\ref{Jcycle}) we have either 
$f(J_j)\cap (\tilde J_k\cup \tilde J_{\hat k})=\emptyset$ such that
$j\not\to\{k,\hat k\}$
or $f(J_j)\supset \tilde J_k$ or  
$\supset \tilde J_{\hat k}$ 
in such a way that $j\to \{k,\hat k\}$
but $f(d_j)\not\in \tilde J_k\cup \tilde J_{\hat k}$. In the first case
\[
\var_{J_k\cup J_{\hat k}}(L(h\cdot 1_J))=0\ .
\]
In the second case $f(d_j)\not\in \tilde J_k\cup\tilde J_{\hat k}$ and 
$\frac{h(x)-h(c_j)}{|f'|}_{|x=c_j}=0$, whence
\beqas
&&\var_{\tilde J_k\cup\tilde J_{\hat k}}(\frac{h-h(c_j)}{|f'|}
\circ f_j^{-1}\cdot 1_{fJ_j})
=
\var_{J_j}(\frac{h-h(c_j)}{|f'|})\\
&=&
\var_{J_j}(\frac{h-h(c_j)}{|f'(c_j)|}\,\frac{|f'(c_j)|}{|f'|})
\le
\var_{J_j}(h-h(c_j))
\cdot\frac{1+\beta}{|f'(c_j)|}+\sup_{J_j}(h-h(c_j))\cdot\beta\\
&\le&
\var_{J_j}(h)\cdot\left(\frac{1+\beta}{|f'(c_j)|}+\beta\right)
\eeqas
by the assumptions on $\beta$ in (\ref{betaassumption}). Hence
\[
\var_{J_k\cup J_{\hat k}}(L(h1_{J_j}))
\le
\var_{J_j}(h)\cdot\left(\frac{1+\beta}{|f'(c_j)|}+\beta
+C\cdot |J_j|\var_{J_j}(h)\right)
\cdot\delta_{j\to\{k,\hat k\}}
\]
Let $\eta_0=\min\{\eta,1\}$. Given $N>0$ and 
choosing $\Z$ sufficiently fine we can make 
$\beta>0$ and $|J_j|$ so small 
that
\[
\frac{1+\beta}{|f'(c_j)|} +\beta
<\frac {1+\frac \ep 2}{|f'(c_j)|} \ ,\quad
C\cdot |J_j|<\frac {\ep\eta_0^{2N}}{|f'(c_j)|}
\quad\mbox{and}\quad
1+\ep < \kappa^{1/N}
\]
for a suitable $\ep>0$ and all $j=1,\ldots,r$. 
(Observe that $r$, the cardinality of $\Z$, 
may depend on the refinement of $\Z$.) Hence 
\be\label{single-estimate}
\var_{J_k\cup J_{\hat k}}(L(h1_{J_j}))
\le
\left(\frac {1+\frac \ep 2}{|f'(c_j)|}
+\frac{\ep\eta_0^{2N}}{|f'(c_j)|}\right)
\cdot\delta_{j\to\{k,\hat k\}}\cdot\var_{J_j}(h)
\ee
for all $j=1,\ldots,r$, and summation over $k$ yields
\beqas
\var_{Y}(L(h1_{J_j}))
&=&
\frac 12 \sum_{k=1}^r\var_{J_k\cup J_{\hat k}}(L(h1_{J_j}))
\le
\left(\frac {1+\frac \ep 2}{|f'(c_j)|}
+\frac{\frac \ep 2}{|f'(c_j)|}\right)
\cdot\var_{J_j}(h)\\
&\le&
\frac{\kappa^{1/N}}{|f'(c_j)|}\cdot\var_{J_j}(h)
\eeqas
for all $j=1,\ldots,r$. This is the special case $n=1$ of the estimate
\be\label{inductive-estimate}
\var_{Y}(L^n(h1_{J_j}))
\le
\frac{\kappa^{n/N}}{\Lambda_n(j)}\cdot\var_{J_j}(h)
\ee
which we now prove for general $n$ inductively:
Suppose (\ref{inductive-estimate}) holds for $n\in\{1,\ldots,N-1\}$. Then
\beqas
\var_{Y}(L^{n+1}(h1_{J_j}))
&=&
\sum_{k=1}^r\var_{Y}(L^n(L(h1_{J_j})1_{J_k}))\\
&\le&
\sum_{k=1}^r\frac{\kappa^{n/N}}{\Lambda_n(k)}\cdot\var_{J_k}(L(h1_{J_j}))\\
&=&
\frac 12 
\sum_{k=1}^r\frac{\kappa^{n/N}}{\Lambda_n(k)}\cdot
\var_{J_k\cup J_{\hat k}}(L(h1_{J_j}))\\
&\le&
\frac 12 \sum_{k=1}^r\frac{\kappa^{n/N}}{\Lambda_n(k)}\cdot
\left(\frac {1+\frac \ep 2}{|f'(c_j)|}\cdot\delta_{j\to\{k,\hat k\}}
+\frac{\ep\eta_0^{2N}}{r|f'(c_j)|}\right)\cdot\var_{J_j}(h)\\
&\le&
\kappa^{n/N}\left(\frac{1+\frac \ep 2}{\Lambda_{n+1}(j)}
+\frac {\ep\eta_0^N} 2\right)\cdot\var_{J_j}(h)\\
&\le&
\kappa^{n/N}\frac{1+ \ep}{\Lambda_{n+1}(j)}
\cdot\var_{J_j}(h)\\
&\le&
\frac{\kappa^{(n+1)/N}}{\Lambda_{n+1}(j)}
\cdot\var_{J_j}(h)
\eeqas
by (\ref{single-estimate}) and (\ref{Lambda-product}).
This proves (\ref{inductive-estimate}) for $n=N$ in particular. Summing 
over all $j$ finally yields the claim of the lemma.
\qed
\begin{corollary}\label{alpha-koro}
If there is some $N$ such that $\Lambda_{N}>1$, 
then there are $C(\Lambda_1,N)>0$ depending continuously on $\Lambda_1$ 
and $\delta>0$ depending only on $f$ and $C$
such that for a partition $\Z$ with $\diam(\Z)<\delta$ and
$\alpha:=\Lambda_N^{-\frac 1{2N}}$ holds
\[
\var_Y(L^nh)
\le
\sum_{j=1}^r\var_Y(L^n(h\cdot 1_{J_j}))
\le
C(\Lambda_1,N)\cdot\alpha^n\cdot\var_Y(h)
\]
for all $n>0$.
\end{corollary}
\proof
Let $\kappa=\Lambda_N^{\frac 12}$. Then the estimate with $C=1$ follows for integer multiples $n$ of $N$ from the previous lemma. To pass from this to
general $n$, apply the previous lemma successively with $N=1$.
\qed
We turn to the finite-dimensional contribution 
$\tilde P_2K$ of $\tilde P_2$: 
Two straightforward inductions yield
\beqa
\tilde P_2^n
&=&
\sum_{k=0}^{n-1}\tilde P_2^{n-k} K L^k\ +\ L^n\quad\mbox{and}
\label{ind1}\\
\tilde P_2^n
&=&
\sum_{k=0}^{n-1}L^k\tilde P_2 K\tilde P_2^{n-k-1}\ +\ L^n\ .
\label{ind2}
\eeqa
Inserting (\ref{ind2}) into (\ref{ind1}) results in
\[
\tilde P_2^n
=
\sum_{k=0}^{n-1}\sum_{l=0}^{n-k-1}L^l\tilde P_2 K\tilde P_2^{n-k-l-1}K L^k
+\sum_{k=0}^{n-1}L^{n-k}KL^k + L^n
\]
whence
{\samepage
\beqa
\tilde P_2^nh
&=&
\sum_{k=0}^{n-1}\sum_{l=0}^{n-k-1}\sum_{i=1}^r\sum_{j=1}^r
L^l(\tilde P_21_{J_j})\cdot(\tilde P_2^{n-k-l-1}1_{J_i})(c_j)
\cdot(L^kh)(c_i)\nonumber\\
&&+\sum_{k=0}^{n-1}\sum_{i=1}^r(L^{n-k}1_{J_i})\cdot(L^kh)(c_i) + L^nh 
\label{Zerlegung}
\eeqa
}
Observe now that for $k\ge 1$
\be\label{Lhochkh}
L^kh=\sum_{i=1}^rL(L^{k-1}h\cdot 1_{J_i})=\sum_{i=1}^r \tilde P_2v_i
\ee
where $v_i=(K-\Id)(L^{k-1}h\cdot 1_{J_i})$ such 
that $\supp(v_i)= J_i$ 
and $v_i(c_i)=0$. 
Therefore
\beqas
\var(\tilde P_2 v_i)
&=&
\var(QP(v_i\cdot1_Y))\\
&\le&
\var(P(v_i\cdot 1_Y))+C\cdot\|P(v_i\cdot 1_Y)\|_1\\
&\le&
\var\left(\frac{v_i}{|f'|}\circ {f_{|J_i}}^{-1}\cdot 1_{fJ_i}\right)
+C\cdot\|v_i\|_1\\
&=&
\var\left(\frac{v_i}{|f'|}\cdot 1_{J_i}\right)+C\cdot\|v_i\|_1\\
&=&
2\,\var_{J_i}\left(\frac{v_i}{|f'|}\right)+C\cdot\|v_i\|_1\,
\quad\quad\mbox{as $v_i(c_i)=0$,} \\
&\le&
\frac 2\eta
\var_{J_i}(v_i)+(C+\const)\cdot\|v_i\|_1\\
&\le&
\frac 2\eta \var_{J_i}(L^{k-1}h)+(C+\const)\cdot\|v_i\|_\infty\cdot|J_i|\\
&\le&
\left(\frac 2\eta+(C+\const)\cdot|J_i|\right)\cdot\var_{J_i}(L^{k-1}h)
\eeqas
with a constant $\const$ depending only on $f$. It follows that for 
sufficiently small $J_i$ (depending on $f$ and $C$)
\beqa\label{Lspectralradius}
2\|L^kh\|_\infty\nonumber
&\le&
\var(L^kh)\\
&\le&
\sum_{i=1}^r \var(\tilde P_2 v_i)
\le
\frac 3\eta\cdot\var_Y(L^{k-1}h)\nonumber\\
&\le&
\frac 3\eta\cdot 
C(\Lambda_1,N)\cdot\alpha^{k-1}\cdot\var(h)\ .
\eeqa
As $|h(c_i)|\le\frac 12 \var(h)$, we can assume that this inequality holds 
for $k=0$, too.
Now (\ref{Zerlegung}) implies
\beqas
&&\var(\tilde P_2^nh)\\
&\le&
\frac 3\eta C(\Lambda_1,N)^2 n\sum_{i=1}^r\left[\sum_{j=1}^r\left(
\var(\tilde P_2 1_{J_j})
\cdot\sum_{s=0}^{n-1}
\alpha^{s-1}
|(\tilde P_2^{n-s-1}1_{J_i})(c_j)|\right)
+\alpha^{n-1}\right]\cdot\var(h)\ ,
\eeqas
and the problem of proving 
exponential convergence of $\var(\tilde P_2^nh)$ to $0$ is
reduced to proving 
exponential convergence of $(\tilde P_2^n1_{J_i})(c_j)$
to $0$. As $\var(\tilde P_21_{J_j})\le\frac 2\eta+(C+\const)\cdot|J_j|$ 
with a constant depending only on $f$, this proves
\begin{proposition}
Suppose there are constants $B,\beta>0$ such that for all $i,j=1,\ldots,r$
\be\label{Linftyass}
|(\tilde P_2^n 1_{J_i})(c_j)|\le B\cdot\beta^n\ .
\ee
Let $\alpha$ be the constant from Corollary \ref{alpha-koro}, 
$\hat\alpha:=\max\{\alpha,\beta\}$. Then there are constants 
$C_2,\delta_2>0$ depending only on $f$, $C$ and $B$ and such that
for a refined partition $\Z$ with $\diam(\Z)<\delta_2$ holds
\[
\var(\tilde P_2^nh)\le C_2\cdot\hat\alpha^n\cdot\var(h)\quad
\mbox{for all $h\in\BV$.}
\]
\end{proposition}
Unfortunately assumption (\ref{Linftyass}) does not follow automatically
from $RW$ or $\ED_{L^1}$. Instead it is implied by the stronger 
assumption $\ED_{L^\infty}$. 
Therefore we discuss the following approach.
By inequality (\ref{Lspectralradius}),
\[
\var(L^kh)
\le
\const\cdot\alpha^k\cdot\var(h)\ .
\]
As $\supp(Lh)\subseteq\tilde Y$ for each $h\in\BV$, $L$ leaves
$\BV(\tilde Y)$ invariant, and $\var=\var_X$ is a norm on $\BV(\tilde Y)$. 
$\tilde K$ is a bounded finite-rank operator on $BV(\tilde Y)$, 
because the 
evaluation of a 
function $h$ at one point is a bounded linear functional.
In particular, also 
$\tilde K_n:=\tilde P_2^n-L^n=(L+\tilde K)^n-L^n$ has finite rank, 
and as
\[
\var((\tilde P_2^n-\tilde K_n)h)=\var(L^nh)
\le
\const\cdot\alpha^n\cdot\var(h)\ ,
\]
$\tilde P_2$ is quasicompact as operator on $\BV(\tilde Y)$. As 
$\tilde P_2$ is at the same time a positive $L^1$-operator, it has a 
nonnegative eigenfuntion $h_0$ to its leading positive eigenvalue $r_0$.
Hence
\be\label{spectral-rad}
r_0^n\int h_0\,dm=\int\tilde P_2^nh_0\,dm\le\const\cdot\omega^n\int h_0\,dm
\ee
for some $\omega\in(0,1)$ by the $\ED_{L^1}$ assumption, which implies
$r_0\le\omega<1$.
\par
Therefore $\var_Y(\tilde P_2^{n-1}h)\le\const\cdot\omega^n\var(h)$,
which is the desired estimate except that the constant, which comes from 
the spectral representation of $\tilde P_2$ cannot be controlled
uniformly in $\ep$. Nevertheless it shows that for each fixed $\ep$ the
perturbed system $f_\ep$ has a smooth invariant measure, and its transition operator satisfies the Lasota-Yorke type inequality.
\par
If $f$ is linear in neighbourhoods of the turning points $c_i$ and if 
$Q_\ep(x,A)=Q(\ep^{-1}(A-x))$ for a fixed probability measure $Q$ (i.e.\ the usual 
scaling behaviour), then the constant 
in (\ref{spectral-rad}) is uniform in $Q_\ep$, because the variation of a function is not changed by a linear change of scale. 
This finishes the
{\bf proof of Theorem \ref{stab4}}. 
\qed
%

%%%%%%%%%%%%%%%%%%%%%%%%
\subsection{Proof of Ulam's conjecture} \label{ulam_s}

Another problem related to stochastic stability is the approximation of chaotic dynamics by finite state Markov chains.
The general idea here is due to Ulam~\cite{Ul}. Consider a partition of 
the phase space $X$ into a finite number of disjoint components 
$\{\Delta_i\}_{i=1}^K, \;  0 < \theta \ep \le |\Delta_i| \le \ep$. 
Then one can compare statistical properties of a map $f$ with those of a Markov chain 
with $K$ states, whose transition probabilities are defined as follows:
$$ p_{ij} = \frac{|f^{-1}\Delta_j \cap \Delta_i|}{|\Delta_i|} $$
Ulam's conjecture states that if one consider a sequence of finite
approximations of this type with $\max |\Delta_i| \to 0$ and $K \to \infty$, 
invariant measures of these Markov chains converge weakly
to a SBR measures of the map $f$. The connection of this problem to the
question of stochastic stability is straightforward, because Ulam's
approximation corresponds to a specific Markov random perturbation where
the transition probability to go from a point $x \in X$ to a 
measurable set $A \subset X$ is equal to 
\be Q_\ep(x,A) := \sum_i \1{\Delta_i}(x) \frac{|A \cap \Delta_i|}{|\Delta_i|},
\label{ulam-tr} \ee
while the corresponding transition operator $Q_\ep: L^1 \to L^1$ is defined 
by
\be Q_\ep h(x) := \sum_i \1{\Delta_i}(x)  \frac1{|\Delta_i|}
      \int_{\Delta_i} h(s) \, ds \label{ulam-tr-op} .\ee
Therefore Ulam's conjecture can be treated in the context 
of random perturbations and it was proved earlier for the case $\la_f>2$ in 
in \cite{Li} (see also discussion of this question in 
\cite{Hu,Ke,Bl5,Bl2}). 

Let $L_\ep^1$ be the finite dimensional linear subspace of $L^1$ 
generated by $\{\frac1{|\Delta_i|} \1{\Delta_i}\}$.
Observe that $Q_\ep L^1= L^1_\ep$.
Consider the transition operator $P_\ep:L_\ep^1 \to L_\ep^1$, defined by 
$$ P_\ep \left({\frac1{|\Delta_i|} \1{\Delta_i}}\right) 
   := \sum_j p_{ij} \frac1{|\Delta_j|} \1{\Delta_j}. $$

\begin{lemma} $P_\ep h = Q_\ep P_f h$ for any function 
$h \in L^1_\ep$. \end{lemma}

\proof It is enough to prove this statement for $h=\1{\Delta_i}$. As
$\int_{\Delta_j}P_f \1{\Delta_i}(s)\,ds=|f^{-1}\Delta_j\cap\Delta_i|$,
we have
$$ Q_\ep P_f \1{\Delta_i}(x) = \sum_j  \frac 1{|\Delta_j|}  \int_{\Delta_j}
     P_f \1{\Delta_i}\,
     \1{\Delta_j}(x) 
   = \sum_j  \frac{|\Delta_i \cap f^{-1}\Delta_j|}{|\Delta_j|} \1{\Delta_j}(x) \CR
   = \sum_j  \frac{|\Delta_i \cap f^{-1}\Delta_j|}{|\Delta_i|} 
     \frac{|\Delta_i|}{|\Delta_j|} \1{\Delta_j}(x) 
   = |\Delta_i| \sum_j p_{ij} \frac1{|\Delta_j|} \1{\Delta_j}(x). $$
Thus
$$ Q_\ep P_f \left( {\frac1{|\Delta_i|} \1{\Delta_i}}(x) \right) 
   = P_\ep \left( {\frac1{|\Delta_i|} \1{\Delta_i}}(x) \right) .$$
\qed

Notice that a related statement was proved in \cite{Li} for the case when 
all intervals $\Delta_i$ have the same length.

Straightforward calculations 
show that the transition probability, defined by  (\ref{ulam-tr}),  is smooth 
and satisfies the variation assumption. Actually it satisfies the stronger
assumption: $\var(Q_\ep h) \le \var(h)$ for any $h \in \BV$.

If $\PTP=\emptyset$, then Ulam's conjecture is a corollary to 
Theorem~\ref{stab2}. However, if $\PTP\ne\emptyset$,
formally the statement of our Theorem~\ref{stab3} can not 
be applied here, because the $RW$-assumption may not be satisfied in 
Ulam's case. Here we prove that under the $H_\infty$-assumption 
Ulam's construction satisfies the $\ED_{L^1}$-condition, such that
{\bf Theorem~\ref{ulam}} follows (just as Theorem~\ref{stab3}) from 
Proposition \ref{RW-ED-lemma}
via Lemma \ref{general-lemma}.

Consider a periodic turning point $c$ with period $N$. We consider 
only small values of $\ep$ and therefore, as in the proof of
Proposition~\ref{RW-ED-lemma}, we may assume here
that the map $f$ is locally linear. Denote 
$$\la:=\min\{\prod_{k=1}^{2N}|f'(f^kx)|:  \;  x,f^Nx \in B_\ep(c)\}; $$
$$\la_k:=|f'(f^{k}c)|, \; \txt{and} \; \Lambda_f:=\max_x|f'x|. $$
Note, that due to (\ref{ulam-tr}) the escape rate (from some interval 
around a turning point) for any point is the same, as for the end-points 
of the corresponding Ulam's interval. Define the neighborhood $Y$ of 
the trajectory of the periodic point $c$ as in the 
Section~\ref{random-walk-assumption}.

\begin{lemma} Let $f \in H_\infty, \; f^N(c)=c$ and the perturbation 
$Q_\ep$ be defined by (\ref{ulam-tr-op}). 
Then there exist constants $\delta>0$ and $n<\infty$ such that
for sufficiently small $\ep>0$ 
$$ \CP\{f_\ep^{2nN}(x) \notin Y\}\ge \delta^{2nN}
\quad\mbox{for each $x\in J$} $$ 
where $J$ is the component of $Y$ containing $c$.
\label{ulam-l} \end{lemma}
\proof In the same way as in the proof of Lemma~\ref{rw-l} consider the 
interval $J$ around the point $c$. In this case it consists of a finite 
number $K_0$ of Ulam intervals. Note that this number is uniformly bounded 
for all $\ep>0$. 

Fix a small positive number 
$\sigma := \theta \ep \gamma/4$, where 
$\gamma:=\frac{\lambda(\lambda-1)}{2N\Lambda_f^{2N}}$. 
Let $\Delta_0$ be the Ulam interval containing the point $c$. 
We set $A:=[c,1] \cap J$ if $|\Delta_0 \cap [c,1]|>|\Delta_0|/2$,  
and $A:=[0,c] \cap J$ otherwise. Define intervals $A_n$ inductively by:
$$ A_0:=\Delta_0 \cap A $$
$$ A_{n+1} := f^{n+1} A \cap \tilde A_{n+1},\mbox{ where }
   \tilde A_{n+1}=\left( \cup_{i: |\Delta_i \cap fA_n| > \sigma} \Delta_i \right) .
$$
Let $n_0=\max\{n: A_{n}\subseteq Y\}$. 
Then $\tilde A_n\subseteq Y$ for $n=0,\ldots,n_0$, and 
it follows by induction that
\begin{itemize}
\item
$A_n$ is an interval adjacent to a point from the orbit of $c$ for $n=0,\ldots,n_0+1$ and
\item
$A_n$ is a union of complete Ulam intervals for $n=0,\ldots,n_0$.
\end{itemize}
Hence
$$ |A_{n+1}| \ge |fA_n| - \sigma \quad\mbox{for }n=0,\ldots,n_0
\label{recursive-estimate}\ ,$$
and a bit more complex calculation yields the following lower bound
\be 
P_\ep \1{A_n}(x) \ge \frac{\gamma}{8\lambda_n}\ge\frac{\gamma}{8\Lambda_f} > 0 \label{inequlam}
\quad\mbox{for any }x\in A_{n+1}. \ee
Indeed, for any Ulam interval $\Delta$ with 
$\Delta \cap A_{n+1} \ne \emptyset$ there exists an Ulam interval $\Delta'$ such that
$\Delta' \cap A_n \ne \emptyset$ and 
$|\Delta \cap f\Delta'|>\min\{\frac\sigma 2,\Lambda_f\theta\ep\}=\frac\sigma 2$.
Therefore the transition probability $p_{\Delta',\Delta}$ to go from 
$\Delta'$ to $\Delta$ can be estimated as follows:
$$ p_{\Delta',\Delta} := \frac{|\Delta' \cap f^{-1}\Delta|}{|\Delta'|} 
   \ge \frac 1{\lambda_n}\cdot\frac{|f\Delta' \cap \Delta|}{|\Delta'|}
   \ge \frac{|\Delta|}{|\Delta'|}\cdot\frac{\sigma}{2\la_{n}\theta\ep}
   = \frac{|\Delta|}{|\Delta'|}\cdot\frac{\gamma}{8\lambda_n} ,$$
such that 
$$P_\ep\1{A_n}\ge p_{\Delta',\Delta}\cdot\frac{|\Delta'|}{|\Delta|}\cdot\1{\Delta}
\ge \frac{\gamma}{8\lambda_n}\ .
$$

Iterating the estimate (\ref{recursive-estimate}) we obtain
$$ |A_n| \ge \la_n \la_{n-1} \dots \la_1 \theta\ep/2 - 
             \sigma \left( \la_n \la_{n-1} \dots \la_2  
             + \la_n \la_{n-1} \dots \la_3 + \dots + \la_n \right) ,$$
where the term
$$ z_n := \la_n \la_{n-1} \dots \la_2 
        + \la_n \la_{n-1} \dots \la_3 + \dots + \la_n $$
can be estimated from above as follows. The sequence $\{\la_k\}$ is periodic 
with period $2N$ and the product $\prod_{k=1}^{2N}\la_k=\la>1$. 
Therefore the highest order term  
contributes most, which gives the following estimate:
$$ z_{2kN} \le \la^{k-1} \frac{2N\Lambda_f^{2N}}{(\lambda-1)}. $$
Thus for $2kN\le n_0+1$ holds
$$ |A_{2kN}| \ge \la^{k} \left(\frac{\theta\ep}2 - 
\frac{2N\Lambda_f^{2N}}{\lambda(\lambda-1)}\sigma \right)
             \ge \la^k \frac{\theta\ep}4  ,$$
which gives an exponential expansion rate for lengths of the intervals $A_n$.
On the other hand by (\ref{inequlam})
$$ P_\ep^{2kN} \1{A_0}(x) \ge \left(\frac{\gamma}{8\Lambda_f}\right)^{2kN} $$
for each point $x \in A_{2kN}$. This finishes the proof because 
the last two inequalities provide a uniform estimate on the necessary
number of steps and the probability to leave $Y$ starting from the 
``centre'' interval $\Delta_0$ of Ulam's partition.  \qed

Now the $\ED_{L^1}$-property for the operator $\tilde P_2$ corresponding to the
Ulam $P_\ep$ follows as in Proposition~\ref{RW-ED-lemma} via Lemma~\ref{integral}.

Remark, that if Ulam's perturbations are selfsimilar for all $\ep>0$ (this means 
that the corresponding transition probabilities do not depend on $\ep$
and that $f$ is piecewise linear), then 
one can prove the convergence of invariant measures in a much more simple 
way. Indeed, if the map $f$ is topologically mixing, then the same is true for 
the corresponding finite Markov chains in Ulam's construction. Now, due to the 
fact that the transition probabilities are the same for any $\ep>0$, we deduce 
that rates of convergence are also the same, which gives the desired statement.

Dealing with Ulam's conjecture, we assume that the elements $\Delta_i$ 
of the partition are of comparable size 
($0 < \theta \le |\Delta_i|/|\Delta_j|\le1/\theta$). To show that this assumption 
is necessary, consider the simplest case of a map $f$ with fixed turning  
point $c=f(c)$ and a family $\{\Delta_i^\ep\}_\ep$ of Ulam intervals, 
such that the intervals near the point $c$ (in the interval $J=J(\ep)$) are 
of length $\ep^2$, while the others are of order $\ep$. Then the number of intervals 
$\Delta_i^\ep$ which have nonempty intersection with $J$ is of order $1/\ep$, and for any fixed 
$\delta>0$ the number of steps for a point close to $c$ to leave the interval $J$ 
with the probability at least $\delta$ goes to infinity as $\ep \to 0$.

\smallskip
Note that Ulam's conjecture seems quite general, and actually we do 
not know any counterexample even for multidimensional hyperbolic maps 
or maps with singularities. For example, the conjecture clearly is true 
for nonchaotic maps with stable periodic orbits.

\subsection{The variation condition on $Q$}
In this section we give sufficient conditions for the Lasota-Yorke-type
property (\ref{var-as}) and (\ref{QLYassumption}) of $Q$. Recall the notations from
Section \ref{stochastic-part} on the stochastic part of the dynamics where we discussed the operator $Q$ and its dual $Q^*$, both defined in terms of the
(sub)-Markovian transition kernel $Q(x,A)$. Extending $Q(x,A)$ to all $x\in\rz$
by setting $Q(x,A)=0$ for $x\in\rz\setminus X$, we can assume that $Q$ and $Q^*$ act on $L^1_m(\rz)$ and $L^\infty_m(\rz)$ respectively.
\par
Let $V=X$, $\tilde V=\rz$ or $V=J_k\cup J_{\hat k}$, 
$\tilde V=\tilde J_k\cup\tilde J_{\hat k}$, and denote by $a$ and 
$b$ the left and right endpoints of $\tilde V$ respectively.
Then $V\subseteq(a+\spread(Q),b-\spread(Q))$ such that
\be\label{ab-infty}
\mbox{$Q^*1_{(a,x]}(a)=1$ and $Q^*1_{(a,x]}(b)=0$ for each $x\in V$.}
\ee
\begin{proposition}
Suppose $Q$ can be decomposed as a sum of linear operators $Q=R+S$ 
in such a way that
\[
\mbox{
$R1_{\tilde V}=1$ on $V$ and $R1_{W}=0$ on $V$ if $W\cap\tilde V=\emptyset$,}
\]
\[
\mbox{$\alpha_R:=\sup_t\var_V(R1_{(a,t]})<\infty$,}
\]
\[
\pbox{$Sh(y)=\int h(x)s(x,y)\,m(dx)$ for some kernel $s$ with
$s(x,y)=0$ if $x\not\in\tilde V$, $y\in V$, and}
\]
\[
C_S:=\sup_x\var_y s(x,y)<\infty\ .
\]
Then
\be\label{LYQ}
\var_V(Qh)\le\alpha_R\cdot\var_{\tilde V}(h)+C_S\cdot\|h\cdot 1_{\tilde V}\|_1
\ee
for each $h\in L^1$.
\end{proposition}
\proof
For $\ph\in\Test(V)$ and $t\in X$ let 
$\Phi(t):=\int_a^t R^*\ph'\,dm$. Then 
$\Phi'=R^*\ph'$ and
\[
\Phi(t)=\int R1_{(a,t]}\cdot\ph'\,dm\ ,
\]
whence
$\Phi(t)=0$ for $t\le a$, and for $t\ge b$ holds:
\[
\Phi(t)=\int R1_{(a,t]}\,\ph'\,dm=
\int_V(R1_{\tilde V}+R1_{(a,t]\setminus\tilde V})\cdot\ph'\,dm=
\int_V\ph'\,dm=\ph(b)-\ph(a)
=0\ ,
\]
because $\supp(\phi')\subseteq V$.
Furthermore
\[
|\Phi(t)|\le\var_V(R1_{(a,t]})\le\alpha_R\ .
\]
Therefore
$\alpha_R^{-1}\Phi\in\Test(\tilde V)$ and
\beqas
\int\ph'\,Qh\,dm
&=&
\int \ph' Rh\,dm + \int \ph' Sh\,dm\\
&=&
\int R^*\ph'\,h\,dm + \int\int\ph'(y)h(x)s(x,y)\,m(dx)\,m(dy)\\
&=&
\int \Phi'\,h\,dm + 
\int\left[h(x)\,1_{\tilde V}(x)\cdot\int s(x,y)\,\ph'(y)\,m(dy)\right]\,m(dx)\\
&\le&
\alpha_R\cdot\var_{\tilde V}(h)+C_S\cdot\|h\cdot 1_{\tilde V}\|_1\ .
\eeqas
\qed
\newpage
\begin{corollary}
The assumptions of the previous proposition are satisfied in each of the 
following two situations:
\begin{enumerate}
\item
$Q$ is a bistochastic kernel (i.e. $Q1=1$) and 
$Q1_{(a,x]}(y)$ is decreasing 
as a function of $y$ for each fixed $x$ (i.e. the ``probability'' 
to reach $y$ 
from $(a,x]$ is decreasing in $y$). In this case $R=Q$ and $S=0$ such that
$\alpha_R=1$ and $C_S=0$.
\item
$Q$ has a differentiable transition density $q(x,y)$ and
\be\label{michaelass}
C_q:=\sup_x\int\left|\frac{\partial q}{\partial x}(x,y)+
\frac{\partial q}{\partial y}(x,y)\right|\,m(dy)<\infty\ .
\ee
In this case $\alpha_R=1$ and $C_S=C_q$.
\end{enumerate}         
\end{corollary}
\proof
\begin{enumerate}
\item
As $Q(x,V)=0$ for $x\not\in\tilde V$, we have $Q1_W=0$ on $V$ if
$W\cap\tilde V=\emptyset$ and $Q1_{\tilde V}=Q1-Q1_{\rz\setminus\tilde V}=1$
on $V$.
For each $t\in\rz$ holds
\[
\var(R1_{(a,t]})=\var(Q1_{(a,t]})
=Q1_{(a,t]}(a)-Q1_{(a,t]}(b)
\le
Q1(a)=1\ .
\]
\item
Let 
$
r(x,y):=-\int_a^y \frac{\partial}{\partial x}q(x,t)\ m(dt)
=-\frac{\partial}{\partial x}Q^*1_{(a,y]}(x)
$, 
define $R$ with this kernel and let 
$s(x,y):=q(x,y)-r(x,y)$. Then $r(x,y)=s(x,y)=0$ if $x\not\in\tilde V$
but $y\in V$, and
\[
R1_{(a,t]}(y)
=
\int_a^t r(x,y)\,m(dx)
=-Q^*1_{(a,y]}(t)+Q^*1_{(a,y]}(a)
=1-Q^*1_{(a,y]}(t)
\]
by (\ref{ab-infty}),
such that $\alpha_R=\sup_t\var(R1_{(a,t]})=\sup_t\var(Q1_{(a,y]})=1$ 
because of the monotonicity of $Q1_{(a,y]}(t)$ as a function of $y$.
Furthermore,
\[
R1_{\tilde V}(y)
=
1-Q^*1_{(a,y]}(b)=1\quad\mbox{for $y\in V$ by (\ref{ab-infty}), and}
\]
\[
R1_W(y)=-\int_Wr(x,y)\,m(dx)=0\quad\mbox{for $y\in V$ if 
$W\cap\tilde V=\emptyset$.}
\]
In order to estimate $\var_y s(x,y)=\var_y(q(x,y)-r(x,y))$ we fix 
$\ph\in\Test(X)$ and consider
\beqas
&&\int\ph'(y)(q(x,y)-r(x,y))\,m(dy)\\
&=&
\int\ph'(y)\left(
q(x,y)+\int_a^y \frac{\partial q}{\partial x}(x,t)\ m(dt)
\right)\,m(dy)\\
&=&
-\int \ph(y)\cdot\left(
\frac{\partial q}{\partial y}(x,y)
+\frac{\partial q}{\partial x}(x,y)
\right)\ m(dy)\\
&\le&
C_q\ .
\eeqas
This estimate shows that $\var_y s(x,y)\le C$ for all $x$.
\qed
\end{enumerate}         
The following corollary allows to apply the reasoning of part 2 of the previous 
one also in cases where $q$ is not differentiable in the strict sense, but where 
e.g. the following condition is satisfied: There is a constant $C>0$ such that for all $\delta>0$ holds
\[
\int|f(x+\delta,y+\delta)-f(x,y)|\,m(dy)\le C\cdot\delta\quad
\mbox{for all }x\ .
\]
\begin{corollary}
If $Q,Q_n$ are (sub)-Markovian operators, $\limn\|(Q-Q_n)h\|_1=0$ 
for all $h$ in a dense subset of $L^1$, and if
$\liminfn C_{q_n}<\infty$ with $C_{q_n}$ as in (\ref{michaelass}), then
\[
\var(Qh)\le\var(h)+\liminfn C_{q_n}\cdot\|h\|_1\ .
\]
\end{corollary}
\proof
$\var(Qh)\le\liminfn\var(Q_nh)$ because $\limn\|Q-Q_n\|_1=0$.
\qed

%--------------------------------------------------------

\end{document}